\documentclass[referee]{raa}            

\usepackage{graphicx,times}             
\usepackage{natbib}
\usepackage{amssymb,amsmath}
\usepackage{amssymb}
\bibpunct{(}{)}{;}{a}{}{,}

\usepackage[a4paper=true,pagebackref=true]{hyperref}
\hypersetup{colorlinks = true, linkcolor = green, anchorcolor = red, citecolor = blue, filecolor = red, pagecolor = red, urlcolor = red}

\begin{document}

   \title{Evaluation of Hadronic Emission in Starburst Galaxies and Star-forming Galaxies
}

   \volnopage{Vol.0 (20xx) No.0, 000--000}      
   \setcounter{page}{1}          

   \author{Yunchuan Xiang
      \inst{1}
   \and  Zejun Jiang
      \inst{1}
   \and Yunyong Tang
       \inst{2}
   }
\institute{$^{1}$Department of Astronomy, Yunnan University, and Key Laboratory of Astroparticle Physics of Yunnan Province, Kunming, 650091, China, xiang{\_}yunchuan@yeah.net, zjjiang@ynu.edu.cn \\
$^{2}$School of Physical Science and Technology, Kunming University, Kunming, 650214, China
}
{\small Accepted~~28-Jul-2021}

\abstract{
In this work, we reanalyzed 11 years of spectral data from the \textit{Fermi} Large Area Telescope (\textit{Fermi}-LAT) of currently observed starburst galaxies (SBGs) and star-forming galaxies (SFGs). We used a one-zone model provided by \textbf{NAIMA} and the hadronic origin to explain the GeV observation data of the SBGs and SFGs.
We found that a protonic distribution of a power-law form with an exponential cutoff can explain the spectra of most SBGs and SFGs. 
However, it cannot explain the spectral hardening components of NGC 1068 and NGC 4945 in the GeV energy band.
Therefore, we considered the two-zone model to well explain these phenomena. 
 We summarized the features of two model parameters,  including the spectral index, cutoff energy, and protons energy budget. 
Similar to the evolution of supernova remnants (SNRs) in the Milky Way, we estimated the protonic acceleration limitation inside the SBGs to be the order of 10$^{2}$ TeV using the one-zone model; this is close to those of SNRs in the Milky Way. 
}

   \authorrunning{Y.C. Xiang, Z.J. Jiang and Y.Y. Tang}            
   \titlerunning{Hadronic Emission of SBGs and SFGs}  

   \maketitle

%
%
\section{Introduction}
\label{sect:intro}

The $\gamma$-ray emission of starburst galaxies (SBGs) and star-forming galaxies (SFGs) can be produced by Bremsstrahlung emission and inverse-Compton scattering of the primary or secondary electrons, as well as the emission from pionic decay, resulting from cosmic-ray (CR) interactions \citep{Abdo2010b,Abdo2010d,Tang2014,Tang2017}. Previous  studies have shown that the pionic decay dominates $\gamma$-ray emissions above 100 MeV for SBGs \citep{Domingo2005,Rephaeli2010}, although leptonic emission is expected to become increasingly important at  low energies.


Thus far, there is no definitive conclusion about the  types of CR propagation effect that play a decisive role in SBGs.
 \citet{Lamastra2017,Lamastra2019} mentioned that the diffusive shock acceleration (DSA) was the primary acceleration mechanism for producing high-energy  particles accelerated in astrophysical shocks.
Moreover, they assumed that DSA was effective in AGN(active galactic nuclei)-driven
shocks for NGC 1068 \citep{Lamastra2019}. 
 However, \citet{Vazza2015,Vazza2016} and \citet{vanWeeren2016} mentioned the difficulties in using    DSA to explain the observed spectra produced by high-energy particles.

For a TeV SBG, NGC 253, \citet{Abdalla2018} believed that  the transport of the cosmic rays (CRs) in NGC 253 was dominated by convection because the hard $\gamma$-ray spectrum of this system attained the observed highest energy band. In addition, its starburst wind with a high velocity implied advection-dominated transport \citep{Abramowski2012}. 
\citet{Peretti2019} believed that the nucleus of an SBG was compact and populated by both gas and unknown sources; they adopted a leaky-box-like model to 
describe the propagation process of the CRs in three SBGs, including M82, NGC 253, and Arp 220. In their  model, they assumed the particle injection of CRs was balanced by the energy losses, advection, and diffusion of the starburst wind.
\citet{Wang2018} built a one-zone model with thick-target materials around SBGs to implement calorimetry and place a firm upper limit on $\gamma$-ray emission from CR interactions for seven recognized SBGs.
 Their model assumed that the CRs of SBGs were accelerated by supernova remnants (SNRs), and all CRs underwent nuclear interactions instead of the escape process. They overlooked the gradient-driven advection and diffusion terms of the CR transport equation in their model.

For a well-known SFG, the Small Magellanic Cloud (SMC), its spectrum with an exponentially cutoff power-law form implied that there was a transition from advection to  diffusion in CR propagation \citep{Lopez2018}. The spectrum of the Large Magellanic Cloud (LMC) was similar to that of the SMC in the GeV energy band \citep[e.g.,][]{Tang2017}. 
This similar spectral feature implies that there may be a similar particle evolution process inside both SFGs.  However, there is currently no convincing evidence to confirm this conjecture.

The uncertainty of particle acceleration inside SBGs and SFGs strongly motivated us to study their high-energy $\gamma$-ray radiation using a one-zone stationary model. 
Here, \textbf{NAIMA}, a spectral research tool \citep[][and references therein]{Zabalza2015}, provides a condition for us to complete this work. 
We used it to analyze the spectra of nine SBGs and four SFGs currently observed in this work. 
The next section presents the results of the \textit{Fermi} Large Area Telescope (\textit{Fermi}-LAT) data analysis, and Section 3 provides the best-fit results of target sources from  one-zone and two-zone models. The discussion and conclusion are presented in Section 4 and 5, respectively.

\section{Data preparation}
\label{sect:Data}
In this analysis, 
the time band of photon events was collected from 2008-08-04 15:43:46 to 2019-08-25 02:54:51. The energy range was from 100 MeV to 500 GeV. 
We used the \textit{Fermi} Science Tools version \textbf{v11r5p3} package\footnote{http://fermi.gsfc.nasa.gov/ssc/data/analysis/software/} provided by the \textit{Fermi} Science Support Center (FSSC). 
We chose the Pass 8 data (evclass = 128 \& evtype = 3) and  the photon events with a zenith angle of 90$^{\circ}$ to minimize the contribution from the Earth's limb. We used \textbf{gtmktime} to obtain high-quality data  with  good time intervals, using the expression of \textbf{(DATA QUAL$>$0) and (LAT CONFIG$==$1)} recommended by the \textit{Fermi} team.  
The instrumental response function \textbf{``P8R3{\_}SOURCE{\_}V2''} was adopted. To reduce the data, we followed the data analysis method provided by the \textit{Fermi} Science Support Center\footnote{//fermi.gsfc.nasa.gov/ssc/data/analysis/}. 
The photon events from a $20^{\circ} \times 20^{\circ}$ region of interest (ROI) were selected. The center coordinates listed in Table \ref{Tab1} were those of target sources from the \textit{Fermi} Large Telescope Fourth Source Catalog \citep[4FGL;][]{4FGL} and SIMBAD\footnote{simbad.u-strasbg.fr}. 
We used the script make4FGLxml.py \footnote{https://fermi.gsfc.nasa.gov/ssc/data/analysis/user/} and 4FGL to generate each source model file.
Owing to a small data sets, we selected a power-law (PL) model as the spectral model of M31 and M33,  based on \citet{Feng2019} and  \citet{Xi2020}.        
For all SBGs, we chose the PL spectral model; for those of SMC and LMC, we selected the log parabola (LOG) model by referring to 4FGL.

Using the binned maximum likelihood method, we fitted the photon events of all 4FGL sources  within the 30$^{\circ}$ range around each target source. Here, the two background templates, including Galactic (\textbf{gll{\_}iem{\_} v07.fits}) and extragalactic diffuse emissions (\textbf{iso{\_}P8R3{\_}SOURCE{\_}V2{\_}V1.txt})\footnote{http://fermi.gsfc.nasa.gov/ssc/data/access/lat/BackgroundModels.html}, 
 were added to each source model file, where their normalizations were set as free parameters. 
Within $5^{\circ}$ of the center of the ROI, the normalization and spectral index of each source were set as free parameters as well.

\subsection{Spectral Energy Distribution}
\label{sect:results}

\begin{figure*}   
\includegraphics[width=50mm]{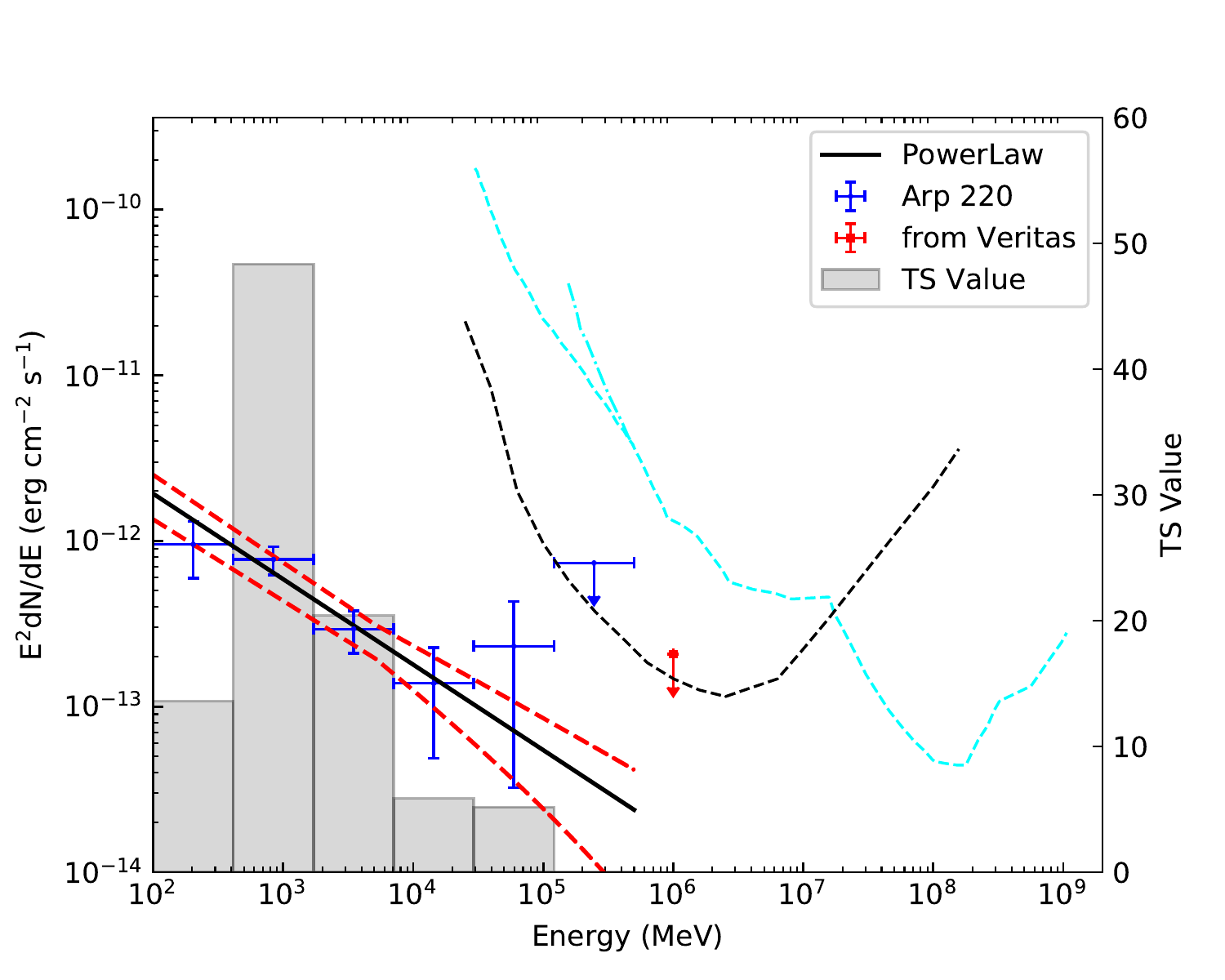}
\includegraphics[width=50mm]{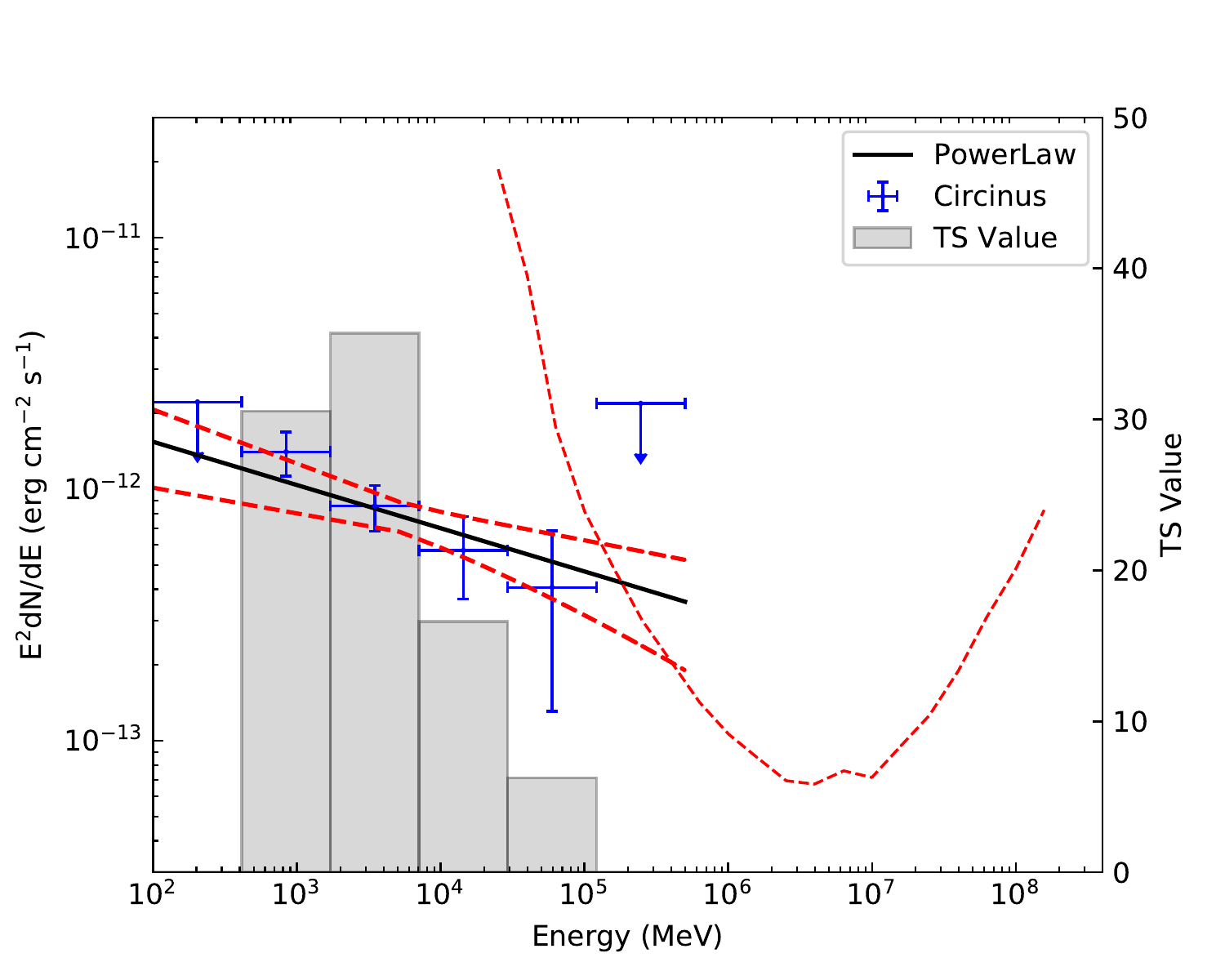}
\includegraphics[width=50mm]{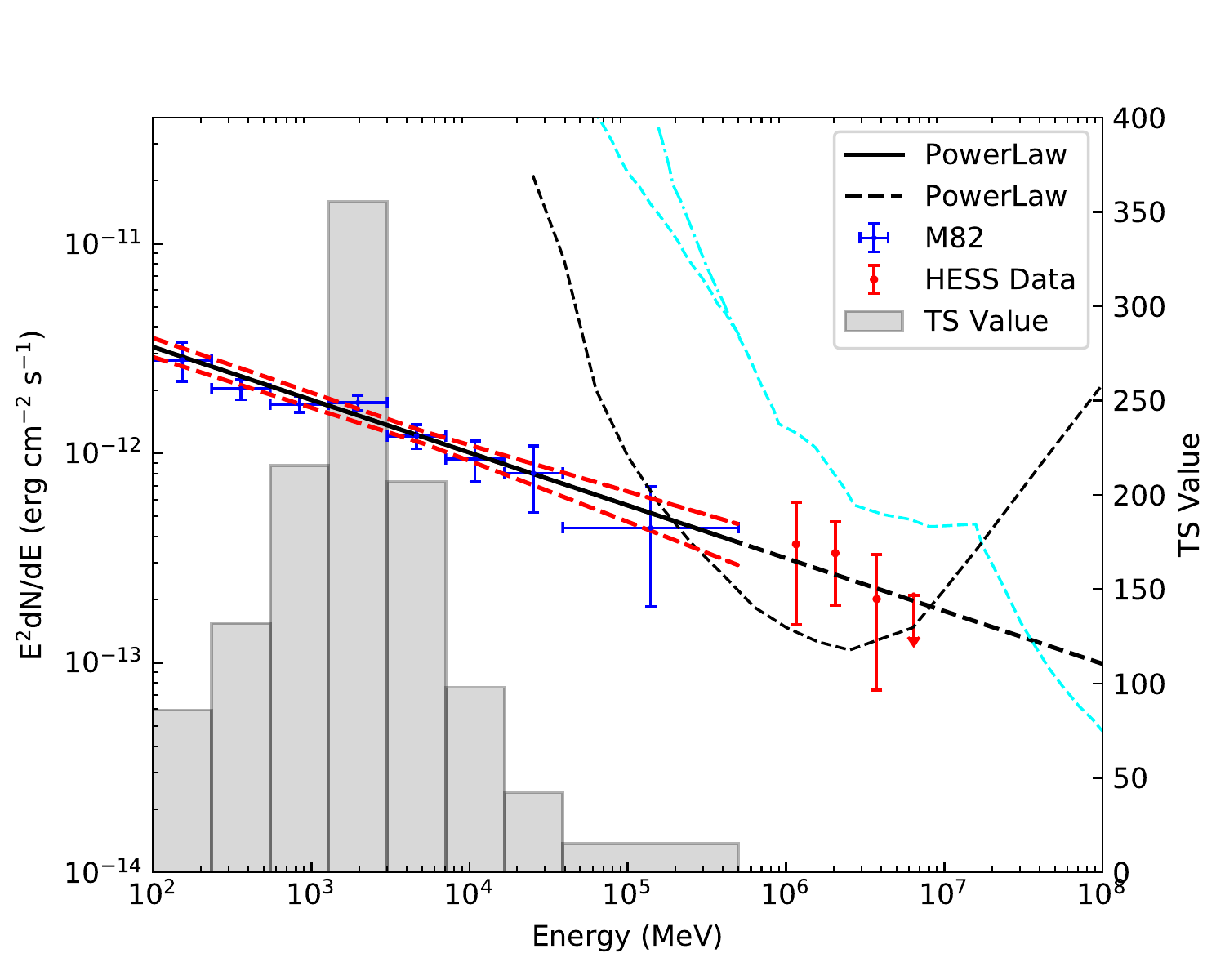}
\includegraphics[width=50mm]{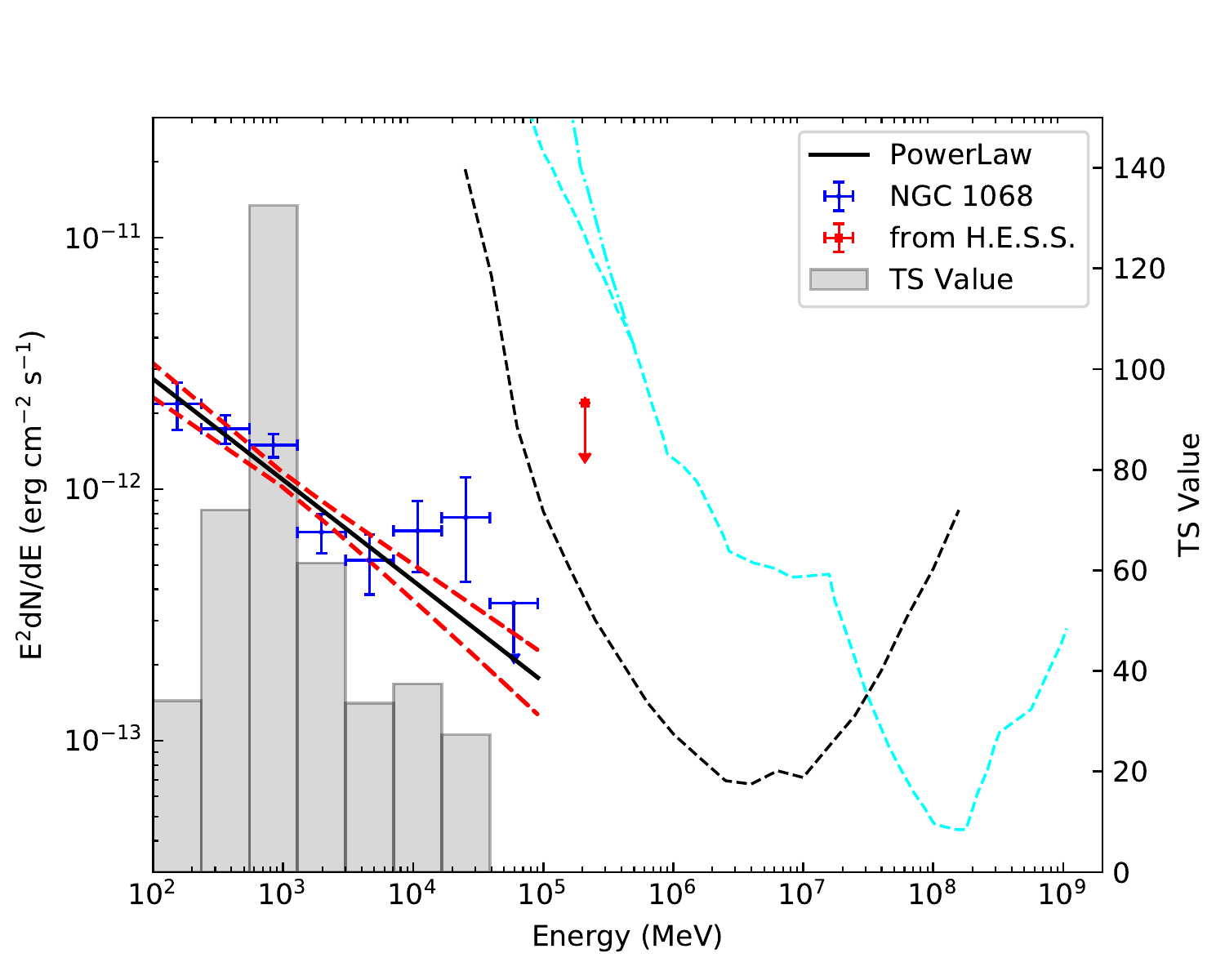}
\includegraphics[width=50mm]{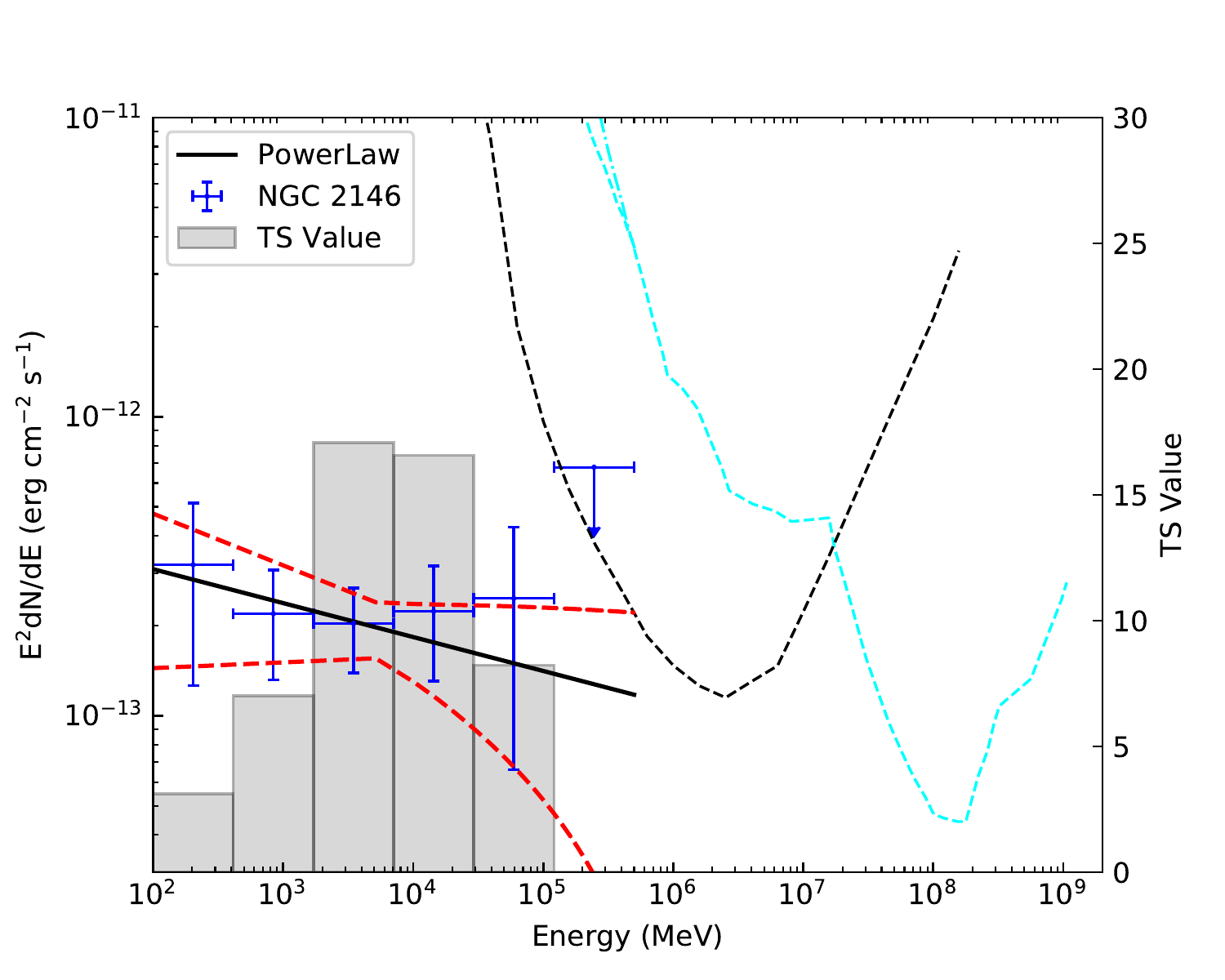}
\includegraphics[width=50mm]{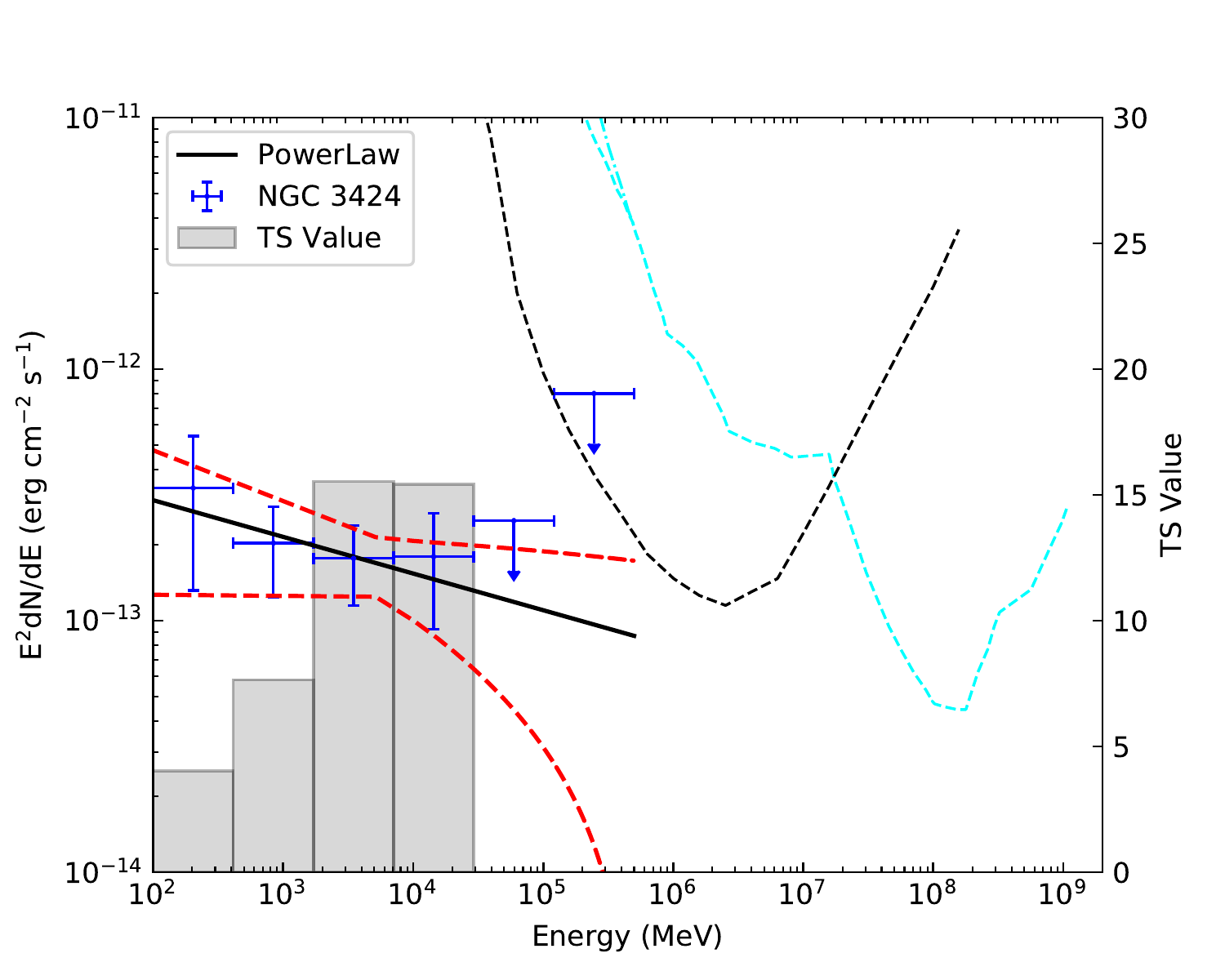}
\includegraphics[width=50mm]{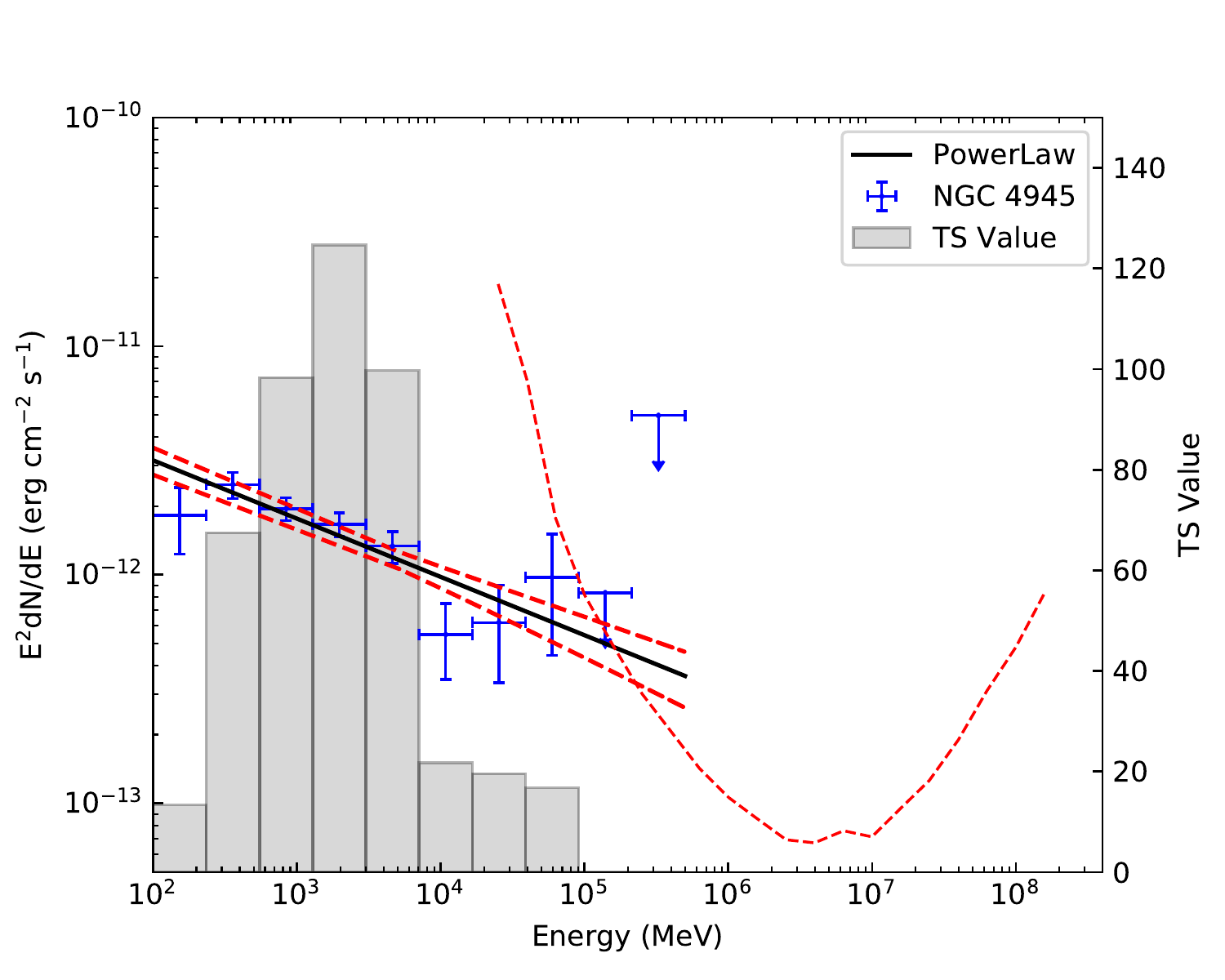}
\includegraphics[width=50mm]{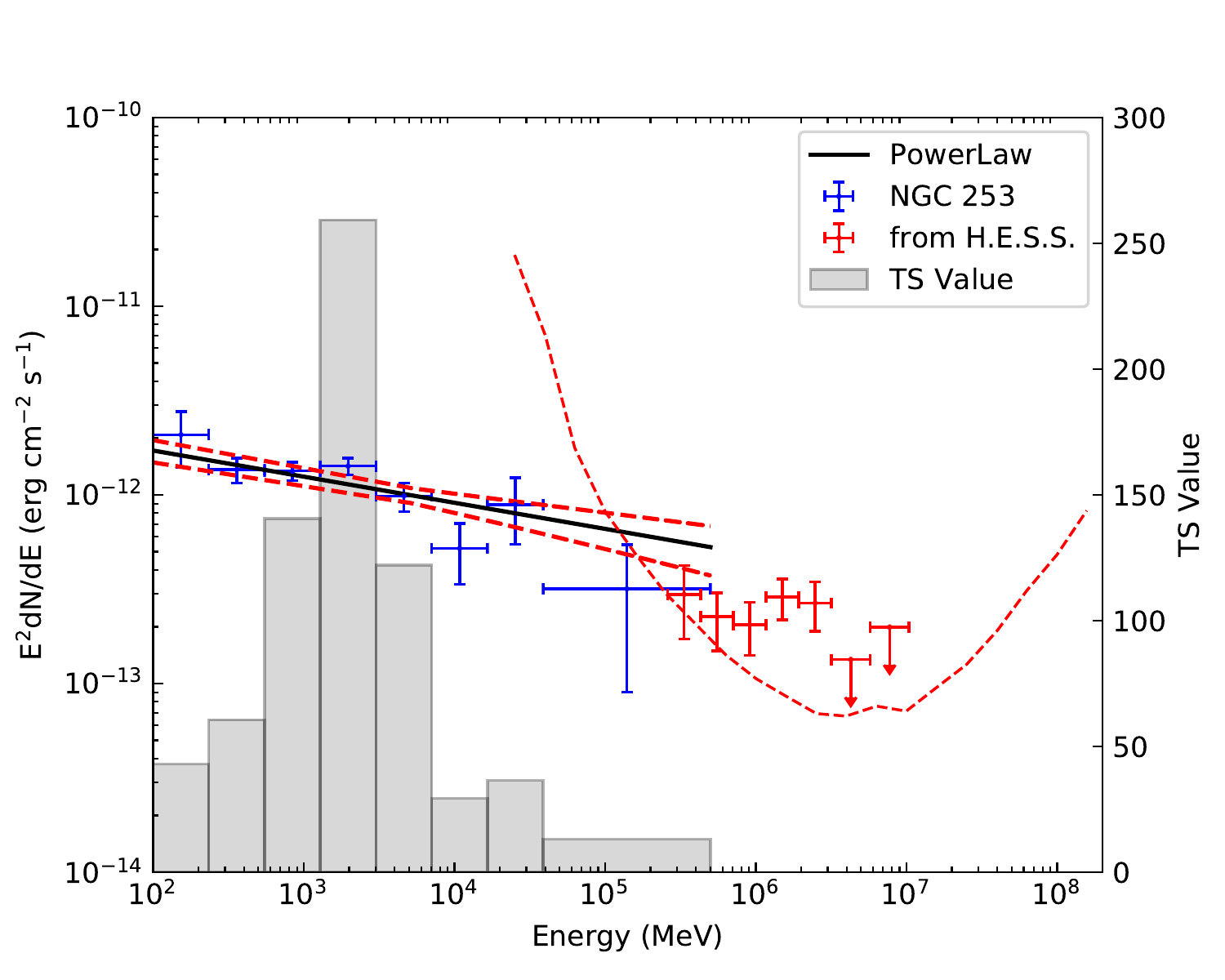}
\includegraphics[width=50mm]{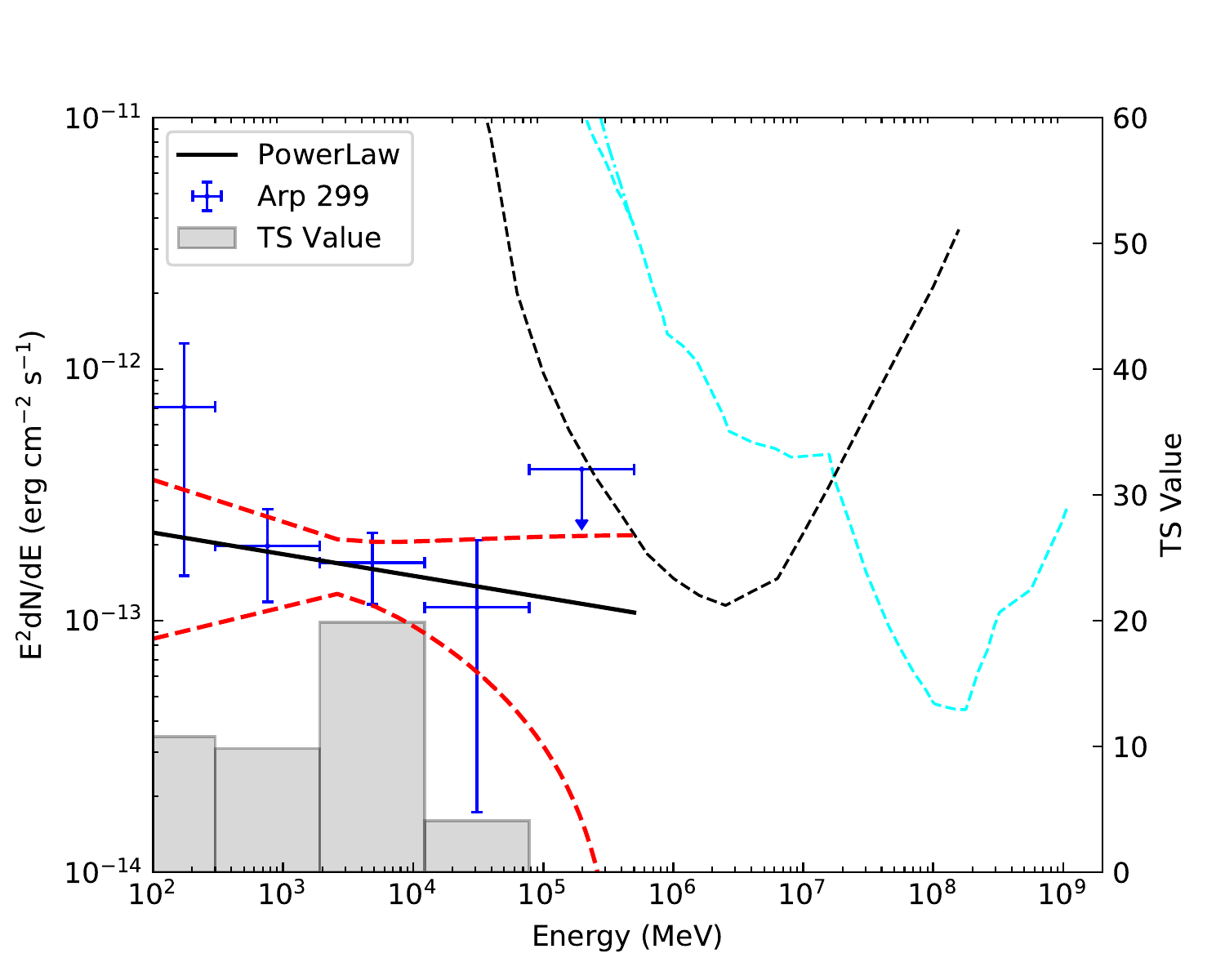}
\includegraphics[width=50mm]{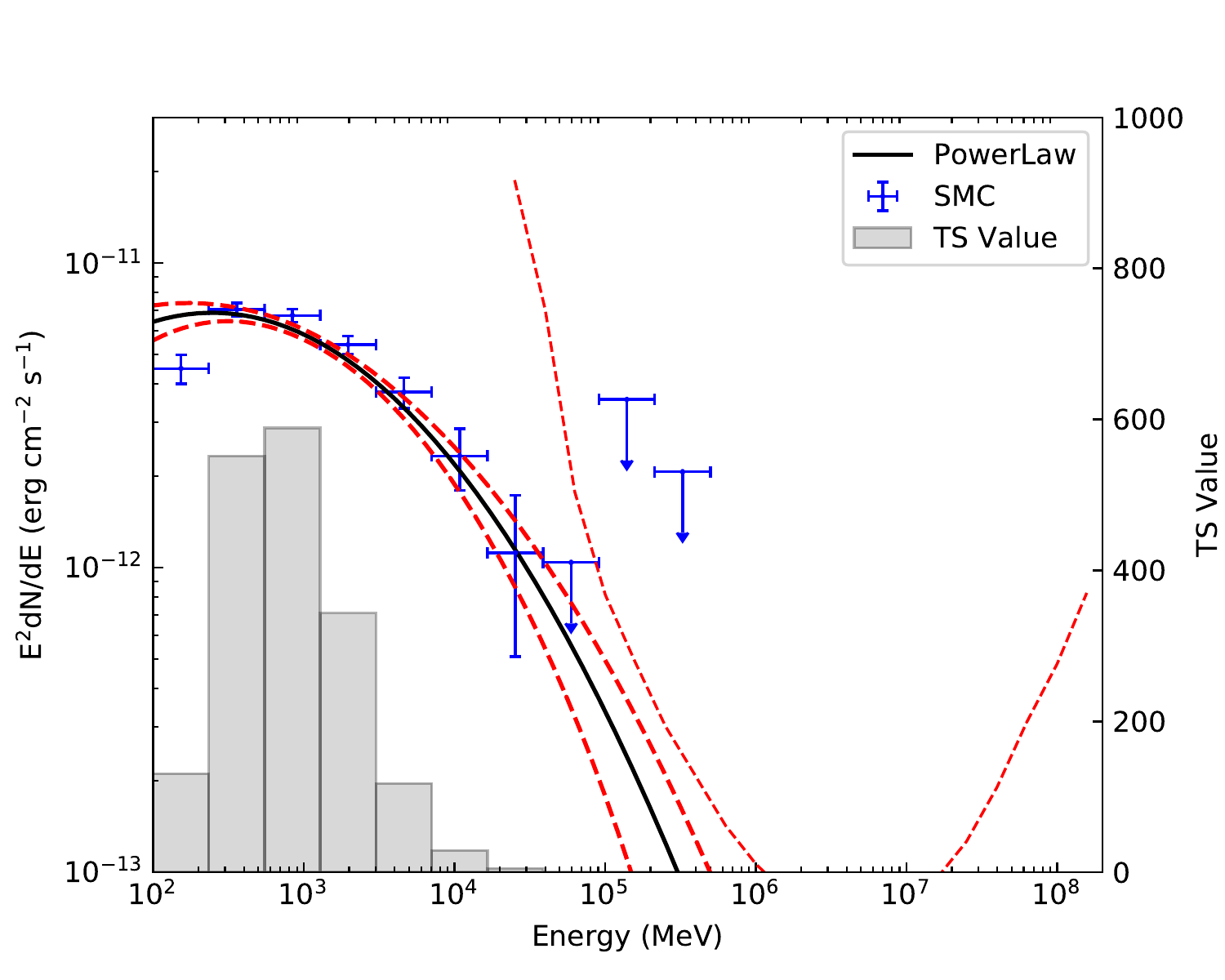}
\includegraphics[width=50mm]{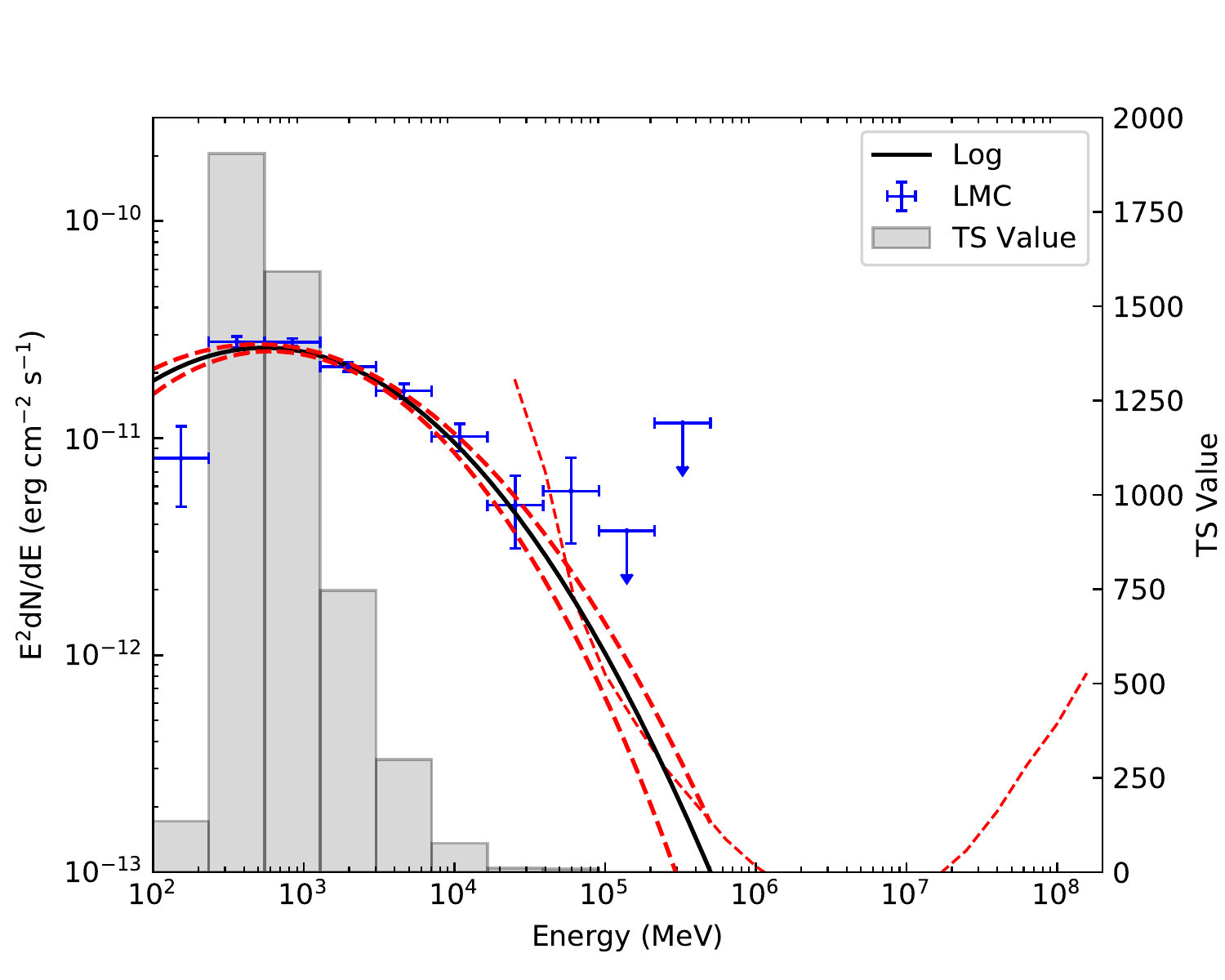}
\includegraphics[width=50mm]{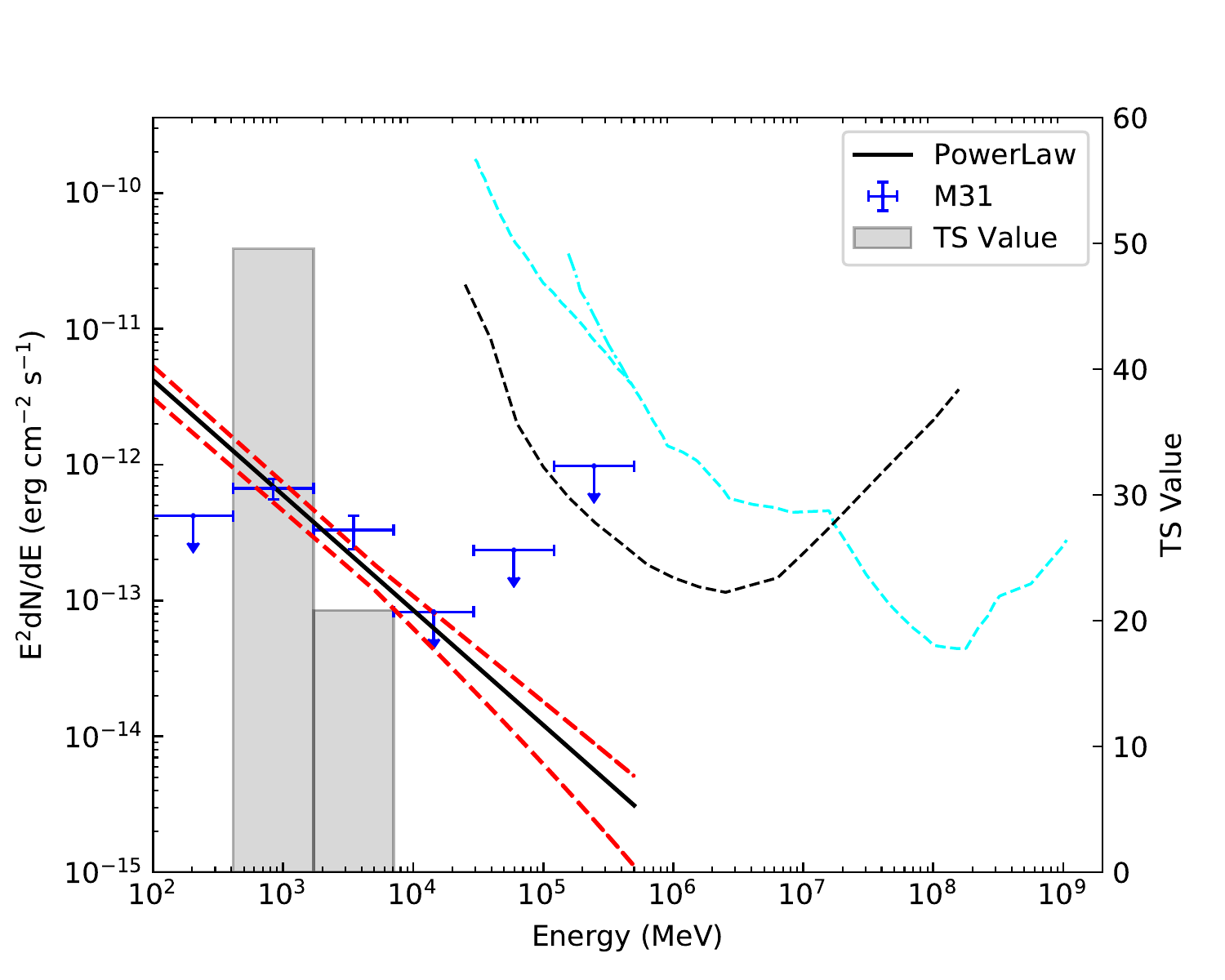}
\includegraphics[width=50mm]{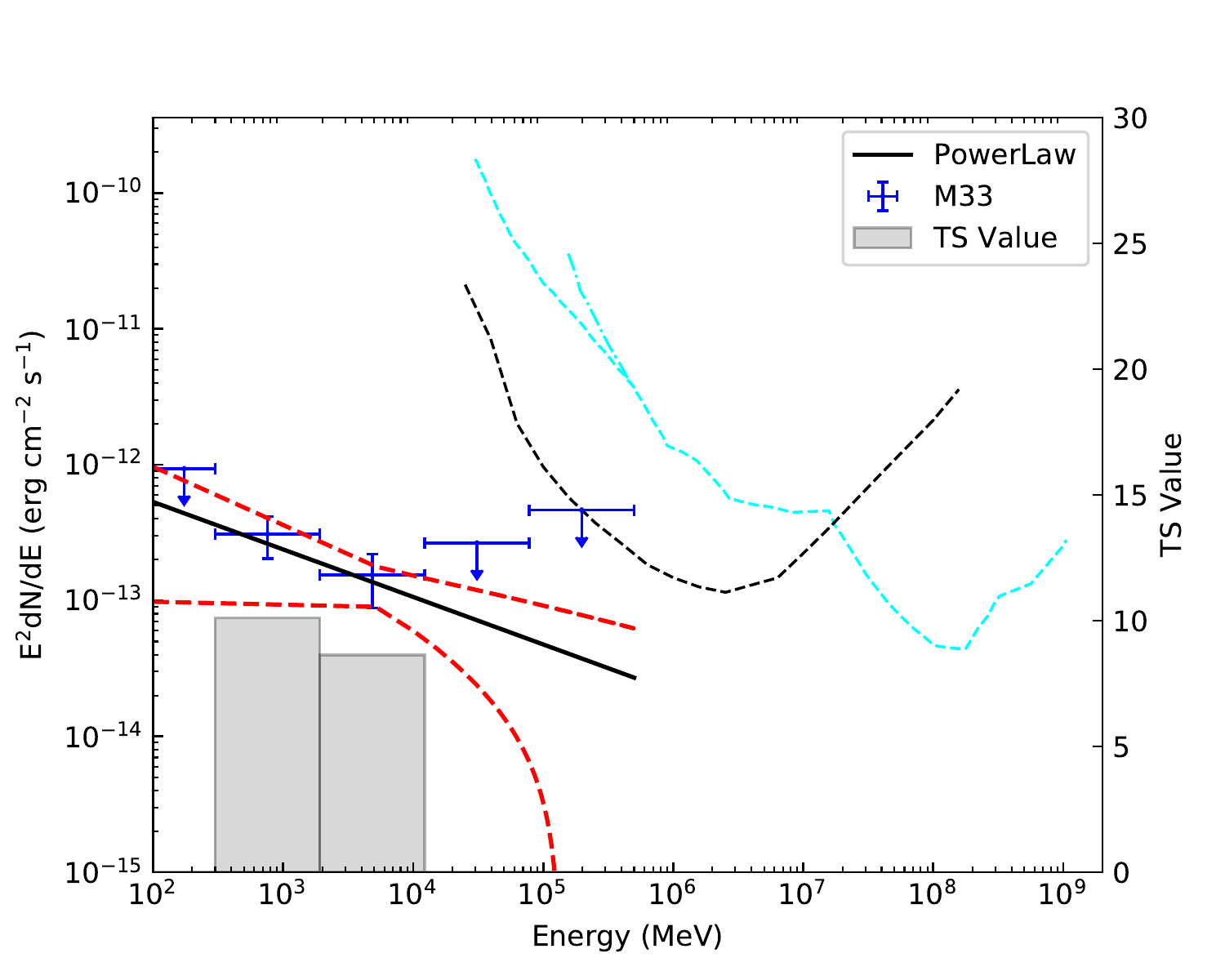}
\caption{The $\gamma$-ray SEDs of SBGs and SFGs. 
The global results and its 1$\sigma$ statistic errors are plotted by the black solid and red dashed lines, respectively. The blue points are the \textit{Fermi}-LAT data from this work. 
The grey shaded area represents the TS value from each energy bin with TS value $>$ 4. 
 The cyan dotted and dot-dashed lines represent the differential sensitivities of LHAASO with different sizes
of photomultiplier tube \citep{Bai2019}. The red and black dashed lines show the sensitivity of CTA-South and CTA-North \citep{CTA2019}, respectively.
For the TeV data, that of NGC 253 is from \citet{Abdalla2018};  
that of M82 is from \citet{Acciari2009}; that of Arp 220 is from \citet{Fleischhack2015}; that of NGC 1068 is from \citet{Aharonian2005}.}
\label{Fig1}
\end{figure*}

To derive the spectral energy distribution (SED) of each source, the energy range from 100 MeV to 500 GeV was divided into six equally spaced log10($E$) energy bins for Circinus, NGC 2146, NGC 3424, M31; for Arp 299 and M 33, we generated five energy bins for a low data statistic, referring to \citet{Xi2020}. 
For the SEDs of other sources, we chose to generate 10 energy bins. 
Because the statistics and TS values from the last three bins were less for NGC 253 and M82, we considered combining the three bins to one, based on  \citet{Abdalla2018}. 
For each bin of all SEDs, we guaranteed that their TS values were greater than 4. 
For the energy bin with a TS value$<$4, the upper limits of the 95\% confidence level were given. 
All SEDs of each source are shown in Figure \ref{Fig1}. 
The global fit results for each source are given in Table \ref{Tab1}. 
We found that the average value of the power-law spectral index of SBGs is approximately 2.23, which is consistent  with the maximum likelihood analysis results of \citet{Ackermann2012a} and \citet{Ajello2020}. 
For each panel of Figure \ref{Fig1}, we provided 
the differential sensitivities of Large High Altitude Air Shower Observatory \citep{Bai2019} and the Cherenkov Telescope Array in the northern/southern hemisphere \citep{CTA2019}, to predict their likely very-high-energy (VHE) emission in the future.

\begin{table*}
\renewcommand\arraystretch{1.5}
\renewcommand\tabcolsep{5.0pt} 
\caption{Spectral Fitting Parameters }
\label{tab:tab1}
\begin{center}
\begin{tabular}{lclclclclclclclclc}
\hline
Source Name & R.A. & Decl. & Spectral Index & Photon Flux (ph cm$^{-2}$s$^{-1}$) & TS value & \\
 \hline
 \multicolumn{6}{c}{The Starburst Galaxy} \\
 \hline
 NGC 253   & 11.90 & -25.29 & 2.14$\pm$0.05   & (9.38$\pm$0.98)$\times 10^{-9}$ & 652.13\\
 M82       & 148.95  & 69.67 &  2.25$\pm$0.04 & (1.60$\pm$0.11)$\times 10^{-8}$ & 1230.72 \\
 NGC 4945  & 196.36  & -49.47 & 2.26$\pm$0.05 & (1.66$\pm$0.13)$\times 10^{-9}$ & 480.51 \\
 Circinus  & 213.29  & -65.33 & 2.17$\pm$0.09 & (5.23$\pm$1.39)$\times 10^{-9}$ & 72.66 \\
 NGC 2146  & 94.53 & 78.33 & 2.11$\pm$0.43 & (1.21$\pm$0.52)$\times 10^{-9}$ & 46.12\\ 
 NGC 1068  & 40.67 & -0.01 & 2.40$\pm$0.06 & (1.22$\pm$0.14)$\times 10^{-8}$& 352.10\\ 
 NGC 3424  & 162.91& 32.89 & 2.15$\pm$0.18 & (1.80$\pm$0.81)$\times 10^{-9}$ &42.01 \\ 
 Arp 299   & 172.07& 58.52 & 2.09$\pm$0.18 & (3.90$\pm$1.34)$\times 10^{-10}$ &33.39 \\
 Arp 220   & 233.70&23.53 & 2.52$\pm$0.12 & (1.94$\pm$1.02)$\times 10^{-9}$ & 35.43 \\
 \hline 
 \multicolumn{6}{c}{The Star-forming Galaxy} \\
 \hline
 M31       & 10.82 & 41.24 &  2.85$\pm$0.50 & (1.49$\pm$0.25)$\times 10^{-9}$ & 72.87\\
 M33       & 23.48 & 30.67 &  2.95$\pm$0.14 & (5.59$\pm$2.07)$\times 10^{-10}$ &19.74  \\
 SMC       & 14.50 & -72.75 & 2.23$\pm$0.03 & (3.98$\pm$0.20)$\times 10^{-8}$ & 1428.44 \\
 LMC       & 80.00 & -68.75 & 2.19$\pm$0.03 & (1.73$\pm$0.06)$\times 10^{-7}$ & 6582.56 \\ 
 \hline
\end{tabular}
\end{center}
\label{Tab1}
\end{table*}

\section{The Analysis of $\gamma$-ray Spectrum} %
\subsection{Model Introduction}
\label{sec:lep} 


The tool, \textbf{NAIMA}, provided a one-zone stationary model and the proton-proton (pp) interaction differential cross-section of PYTHIA 8 \citep{Kafexhiu2014} to fit the spectra of the target sources. This tool employs the Markov chain Monte Carlo method, implemented in the \textbf{emcee} package \citep{Foreman2013}.  We will use \textbf{NAIMA} to complete all the following analyses.

The high-energy radiations in SFGs and SBGs are thought to be caused by CRs colliding with the surrounding interstellar medium (ISM) resulting in the generation of  neutral pions and other products 
\citep{Abdo2010b,Abdo2010d,Peretti2019}. In addition, the neutral pions can decay into high-energy $\gamma$-ray photons \citep{Stecker1971,Dermer1986}; thus, the SBGs and SFGs were considered to be possible $\gamma$-ray sources in the local universe \citep{Dermer1986,Strong1976,Paglione1996,Blom1999,Domingo2005}. 
The $\gamma$-ray radiation of SBGs and SFGs above 100 MeV is widely regarded to be of hadronic origin; whereas pionic decay is regarded as the dominant mechanism of $\gamma$-ray generation inside them \citep[e.g.,][]{Domingo2005,Rephaeli2010,Tang2014,Abdalla2018,Peretti2019,Wang2018}. 
Therefore, we assumed that the high-energy radiation of SBGs and SFGs originated from the collective interaction of internal SNRs and molecular cloud gas inside those starburst regions, and that pionic decay is the dominant radiation mechanism for the GeV $\gamma$-ray emission of the SBGs and SFGs. 
For the absorption of extragalactic background light (EBL), here we consider the EBL model from \citet{Dominguez2011} in all the following analyses.

For SNRs in the Milky Way, the protonic energy distribution of a power-law distribution with an exponential cutoff (ECPL) has been widely used to explain their $\gamma$-ray spectra \citep{Aharonian2006,Xing2016, Xin2019,Xiang2021a}. 
Here, we assumed that the protonic energy distributions of SNRs within  SBGs and SFGs satisfy the same formula,  which is as follows:
\begin{equation}
 N(E) = N_{0}\left (\frac{E}{E_{0}}\right )^{-\alpha}exp\left (-\frac{E}{E_{\rm cutoff}}\right),
\end{equation}
where $N_{0}$ is the amplitude, $E$ is the particle energy, $\alpha$ is the power-law spectral index, and $E_{\rm cutoff}$ is the cutoff energy, and $E_{0}$ = 1 TeV.

\subsection{Model fit}

\begin{figure*}[!h]
\includegraphics[width=50mm]{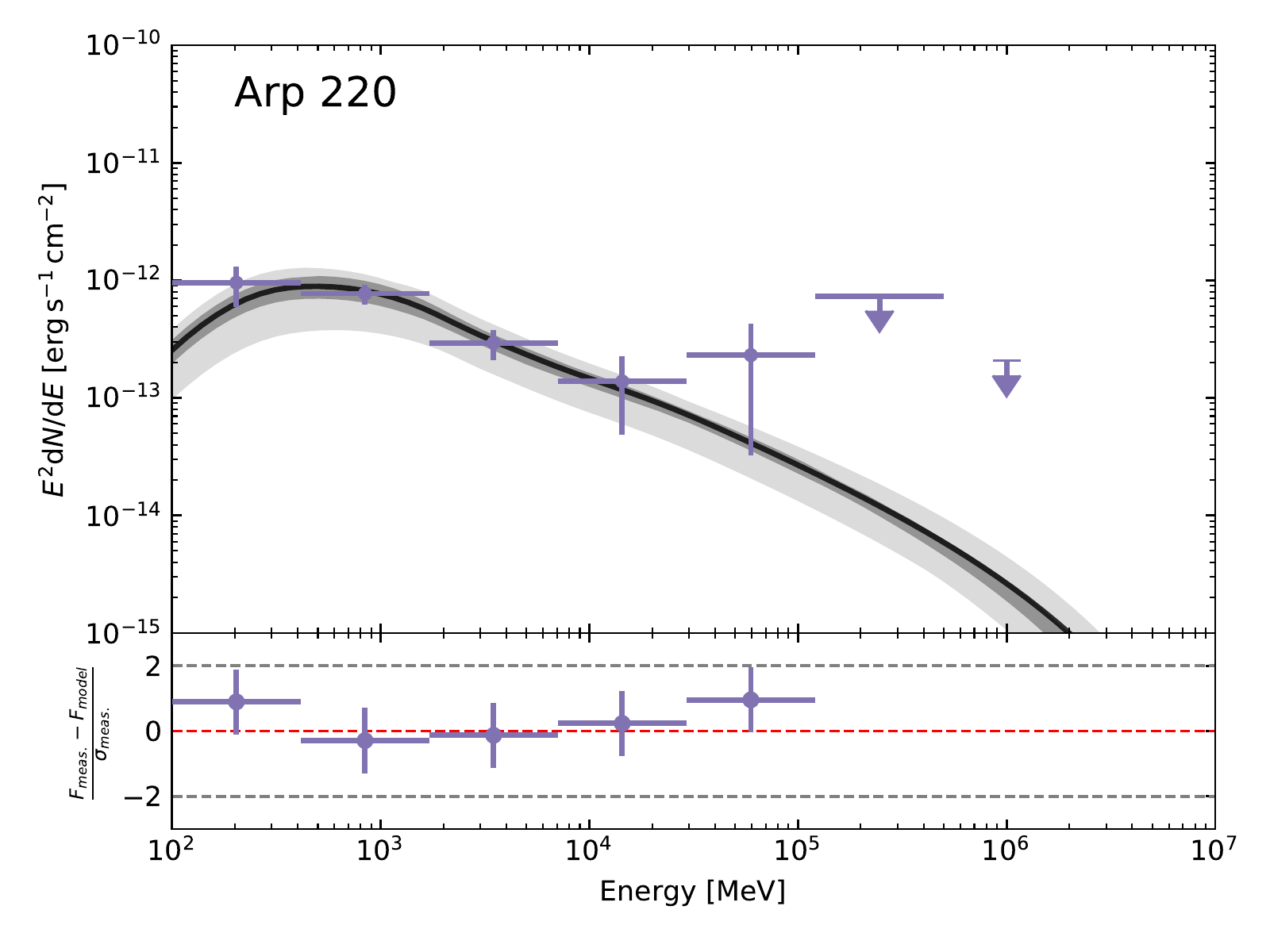}
\includegraphics[width=50mm]{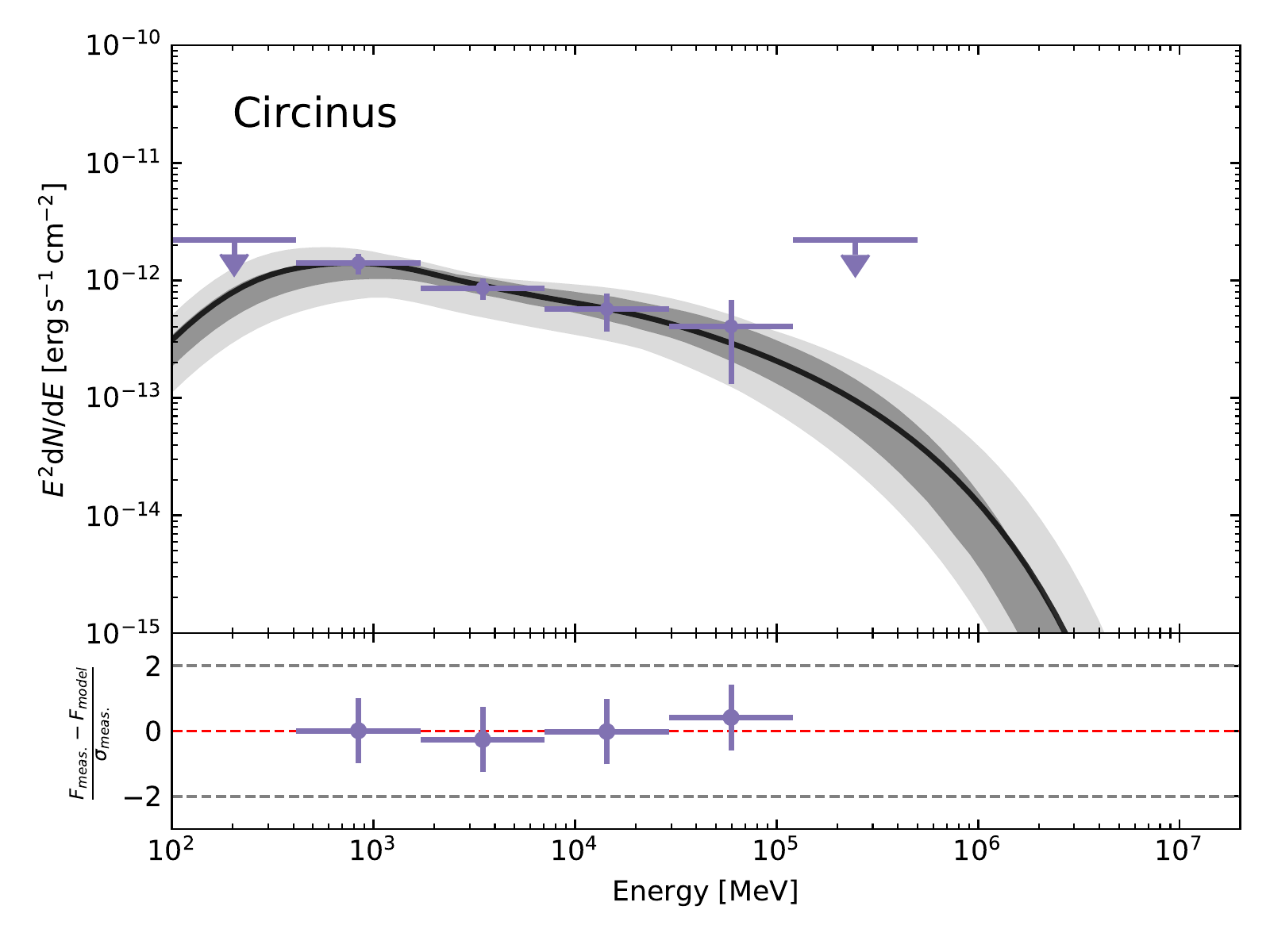}
\includegraphics[width=50mm]{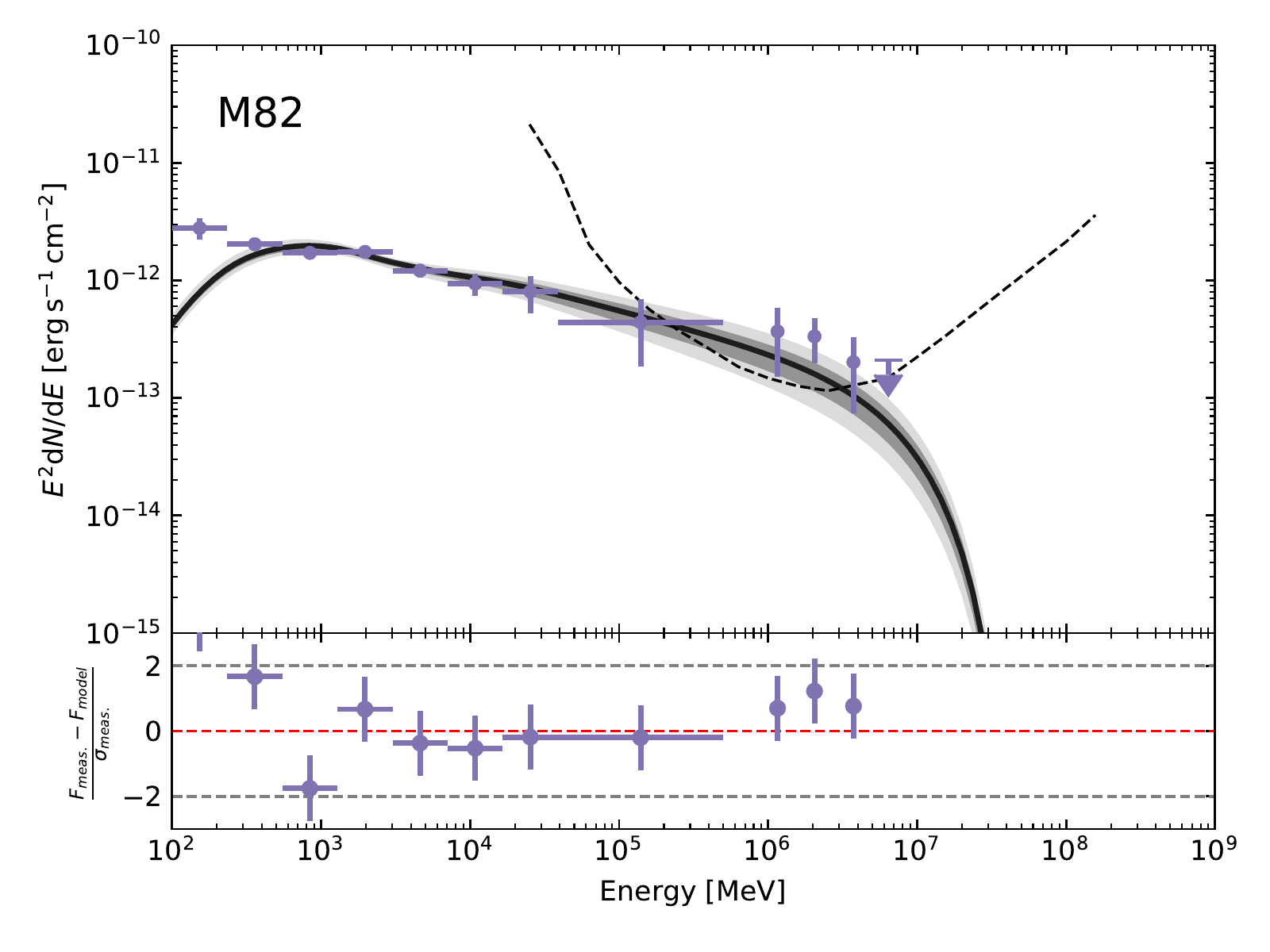}
\includegraphics[width=50mm]{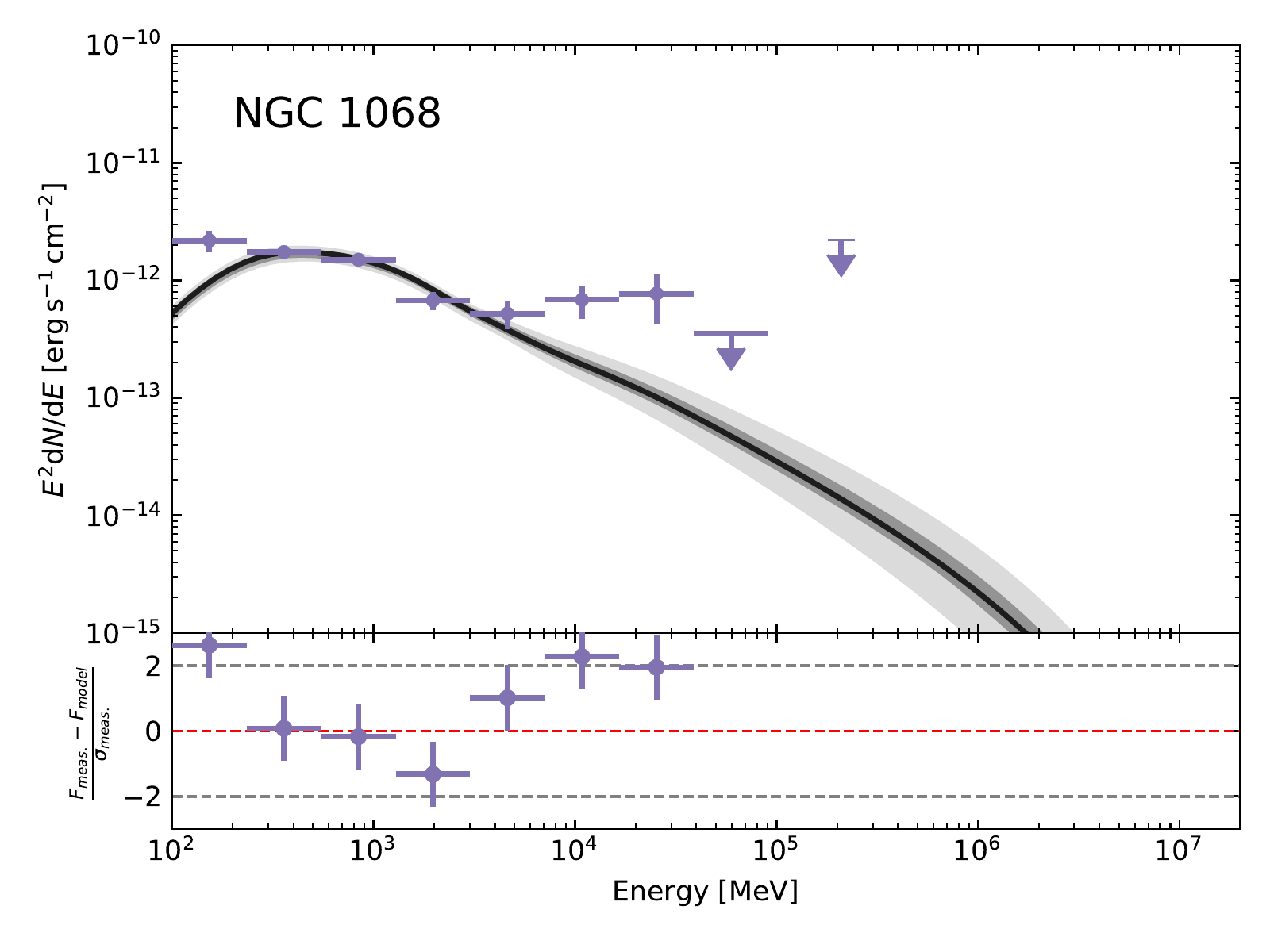}
\includegraphics[width=50mm]{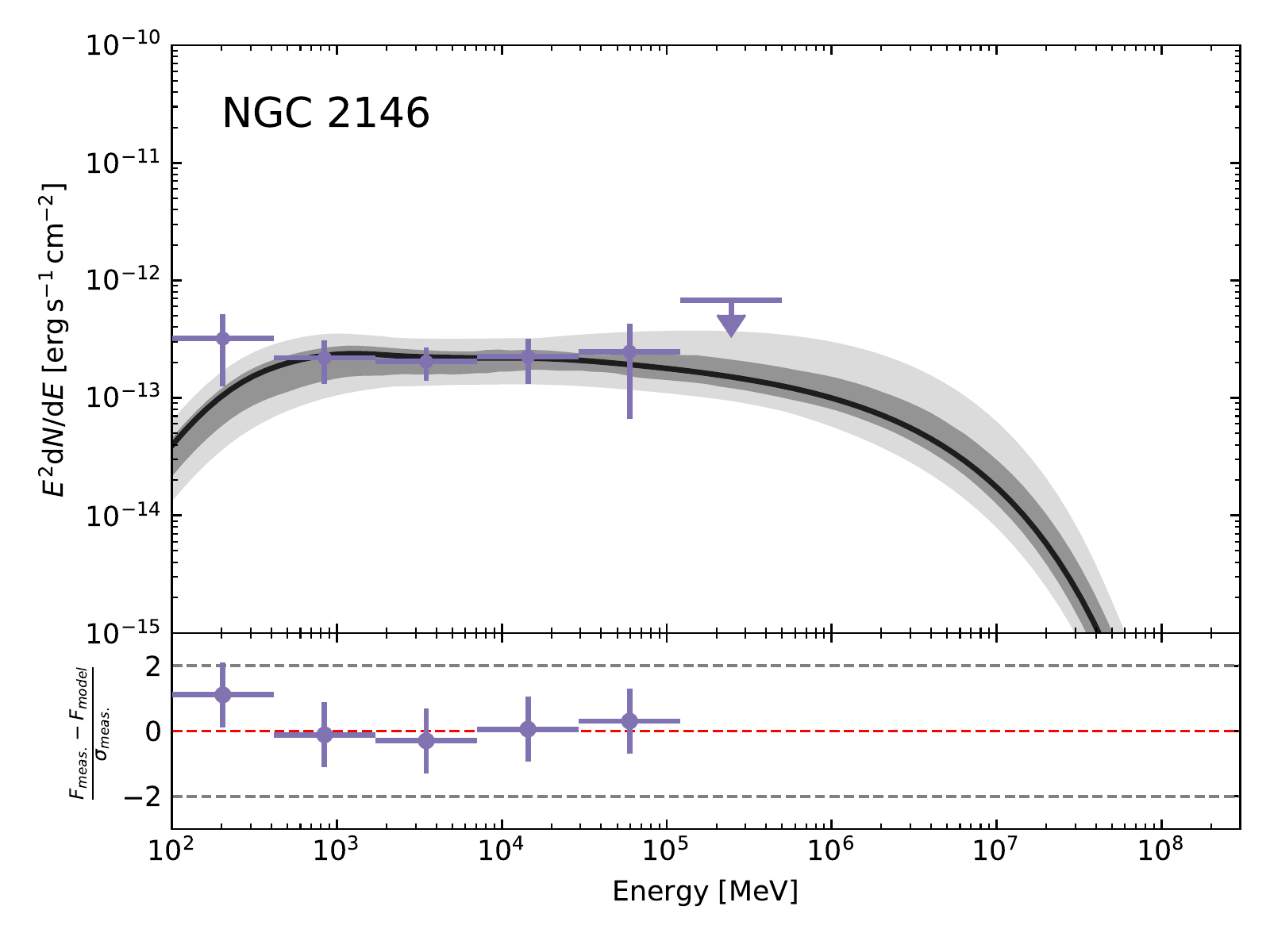}
\includegraphics[width=50mm]{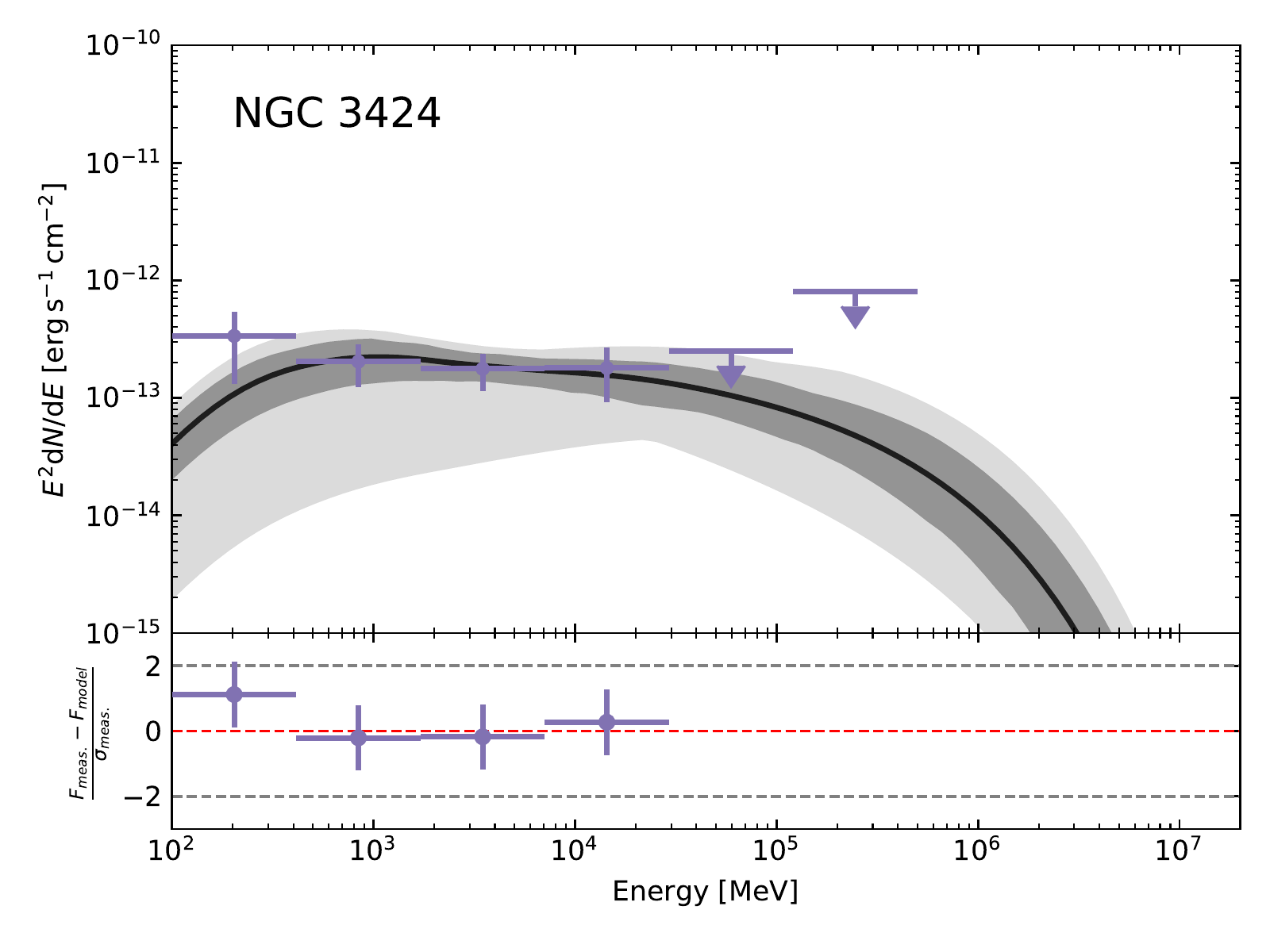}
\includegraphics[width=50mm]{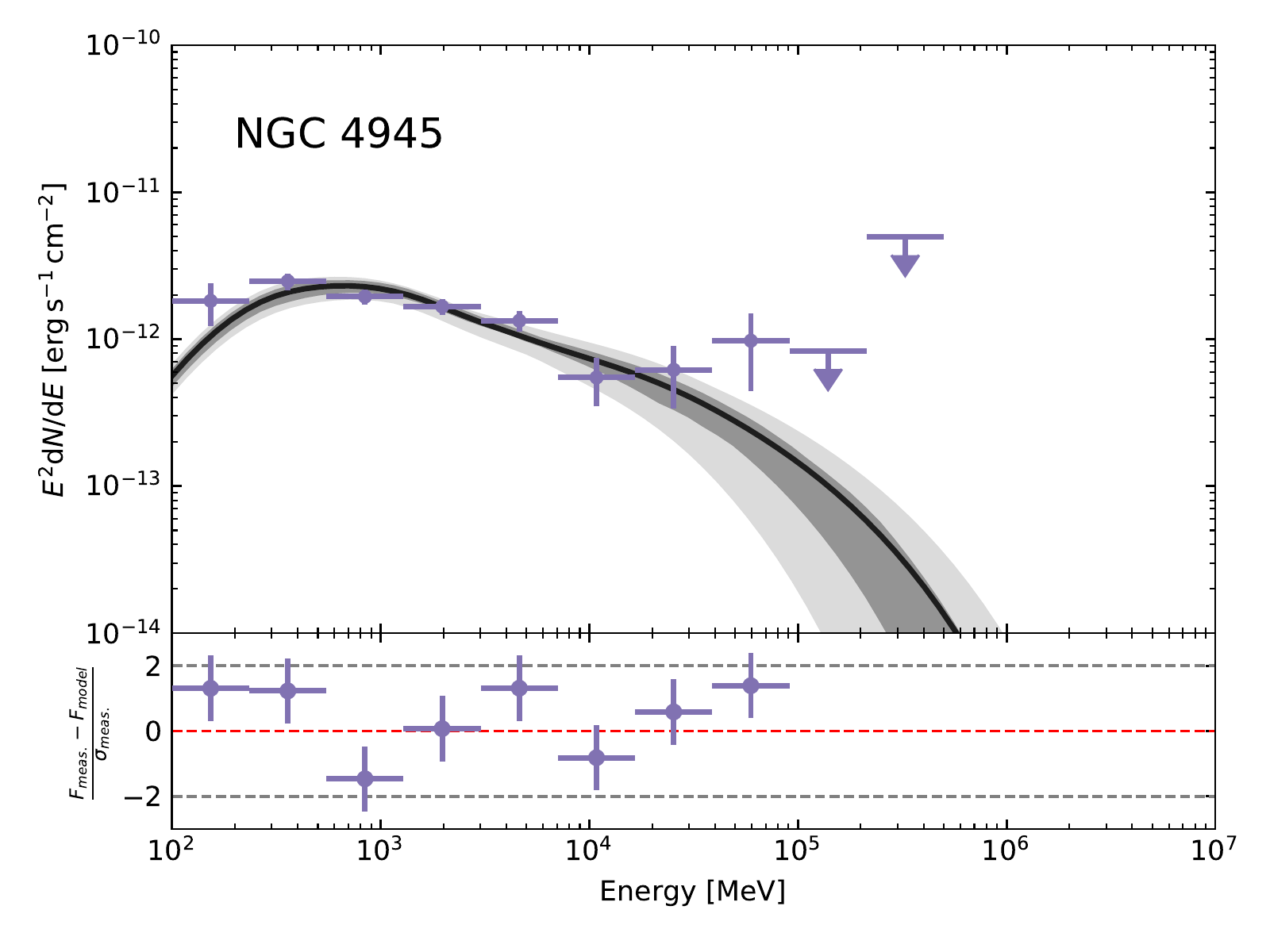}
\includegraphics[width=50mm]{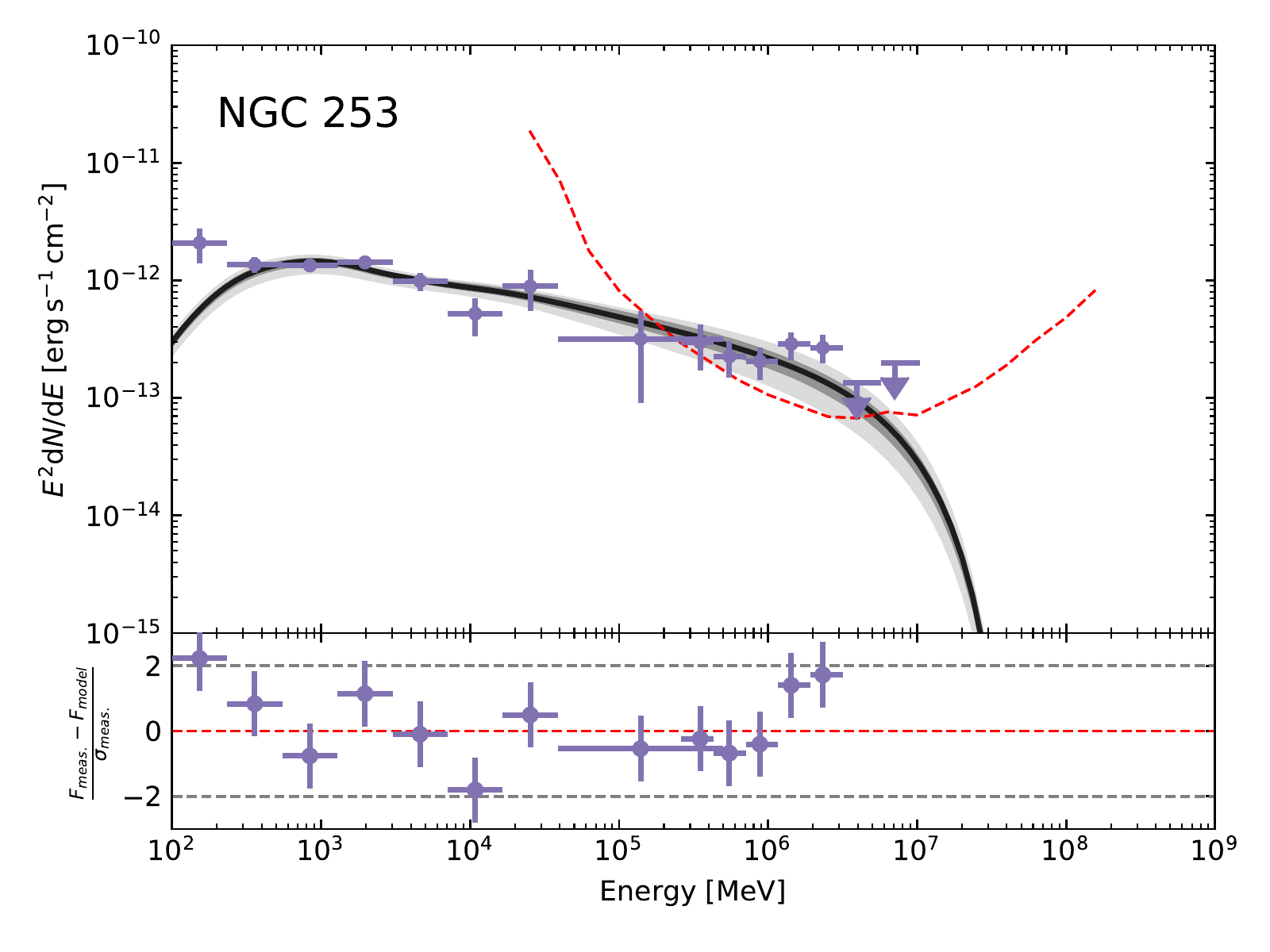}
\includegraphics[width=50mm]{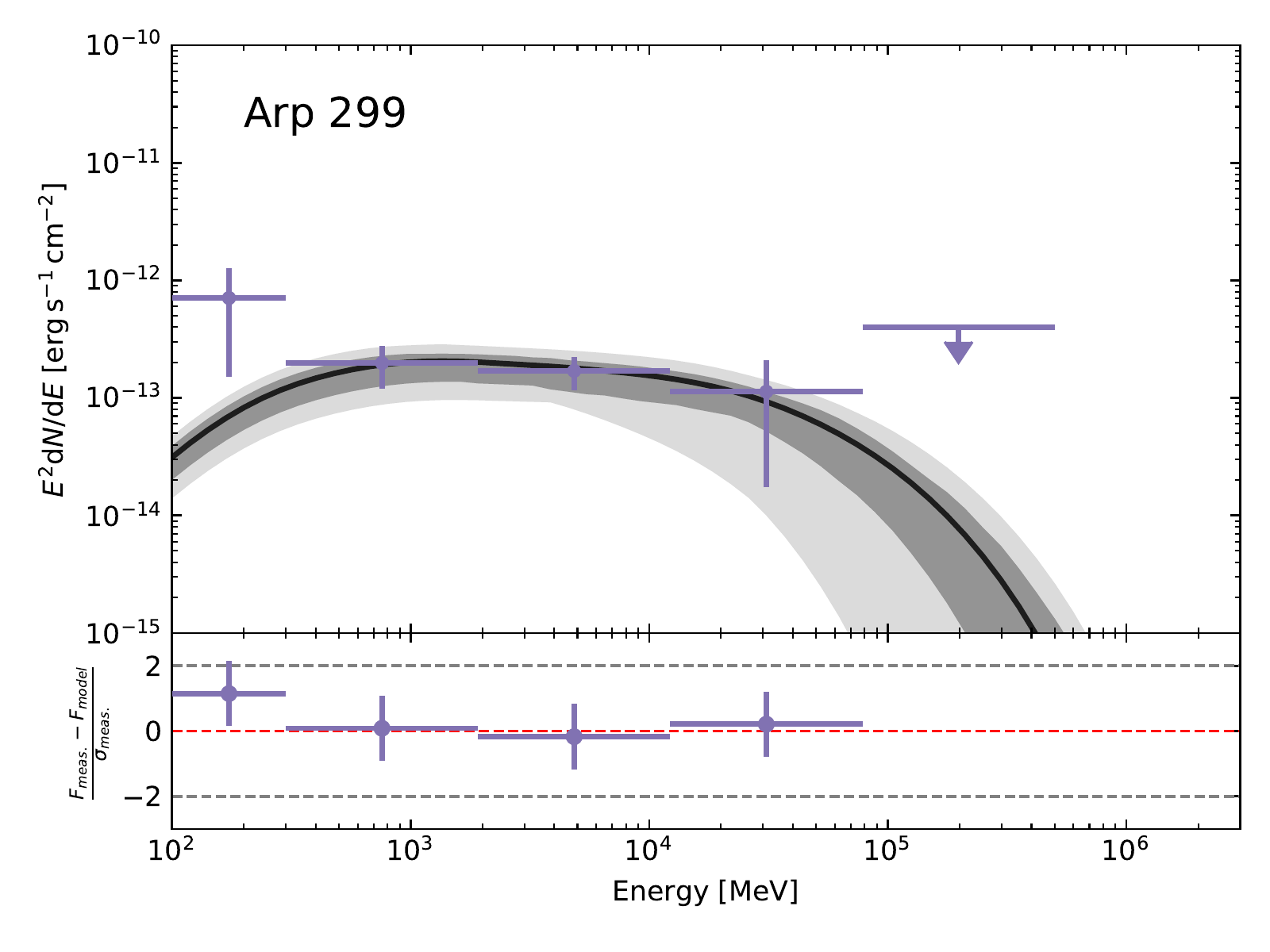}
\includegraphics[width=50mm]{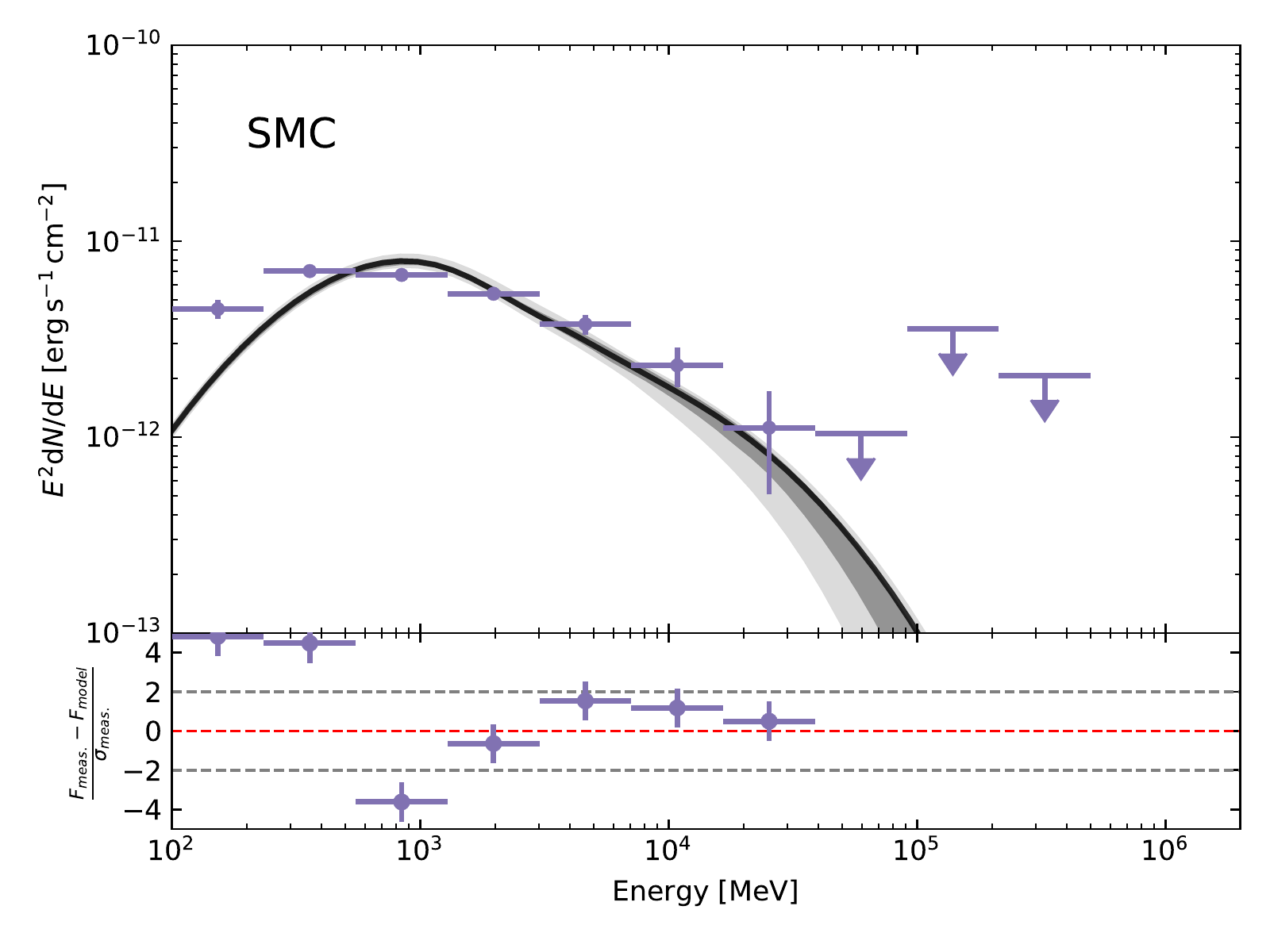}
\includegraphics[width=50mm]{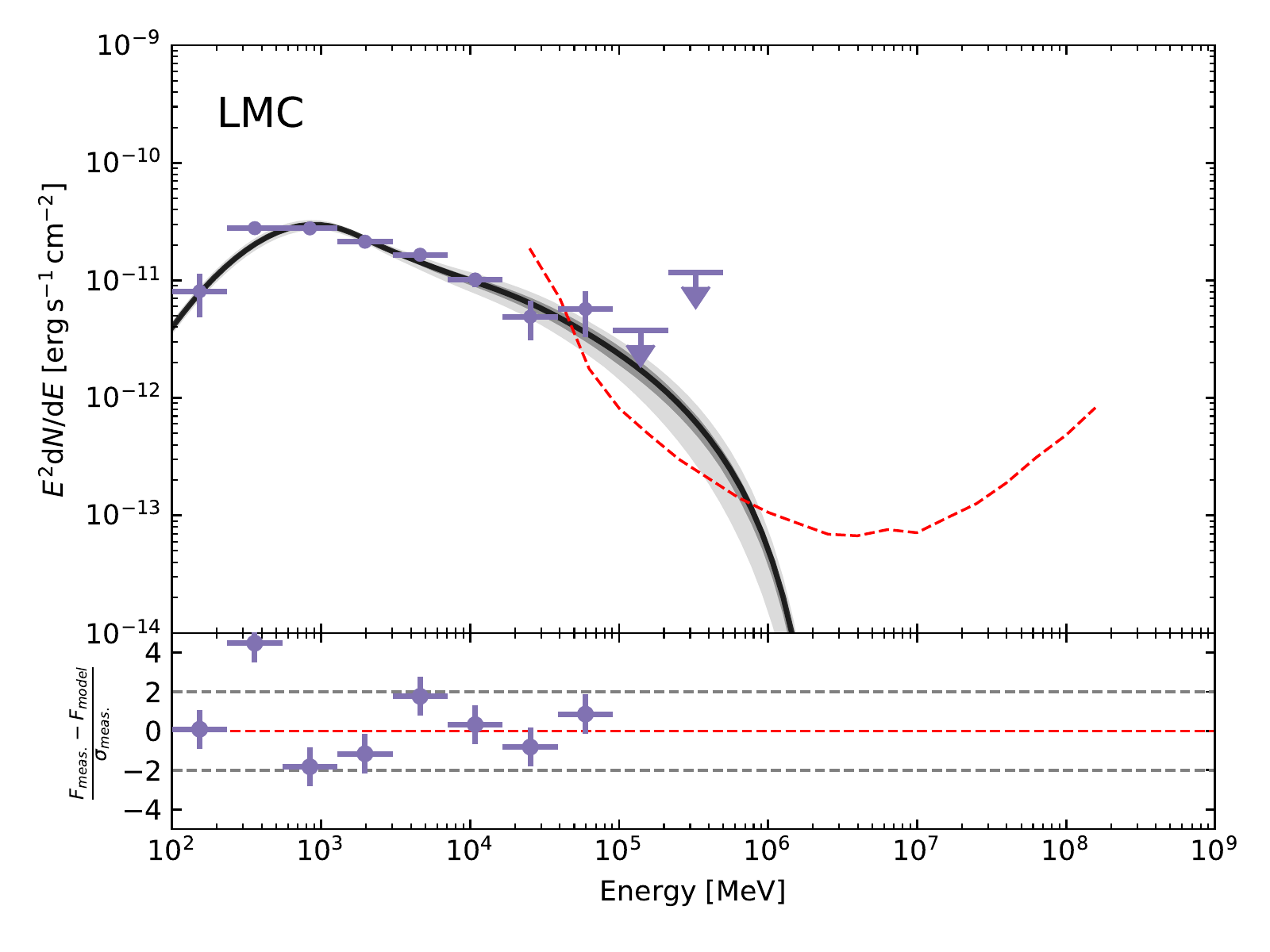}
\includegraphics[width=50mm]{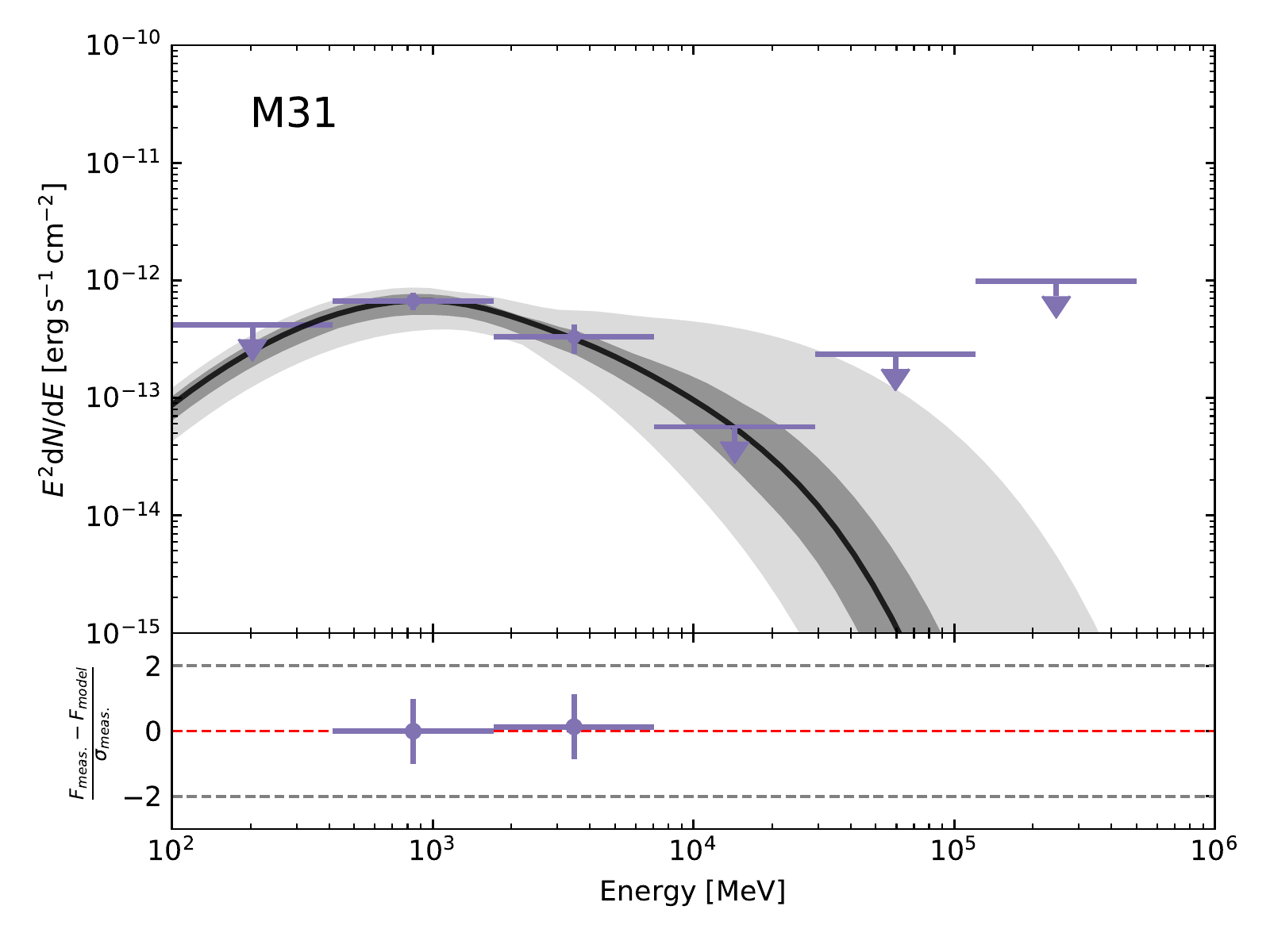}
\includegraphics[width=50mm]{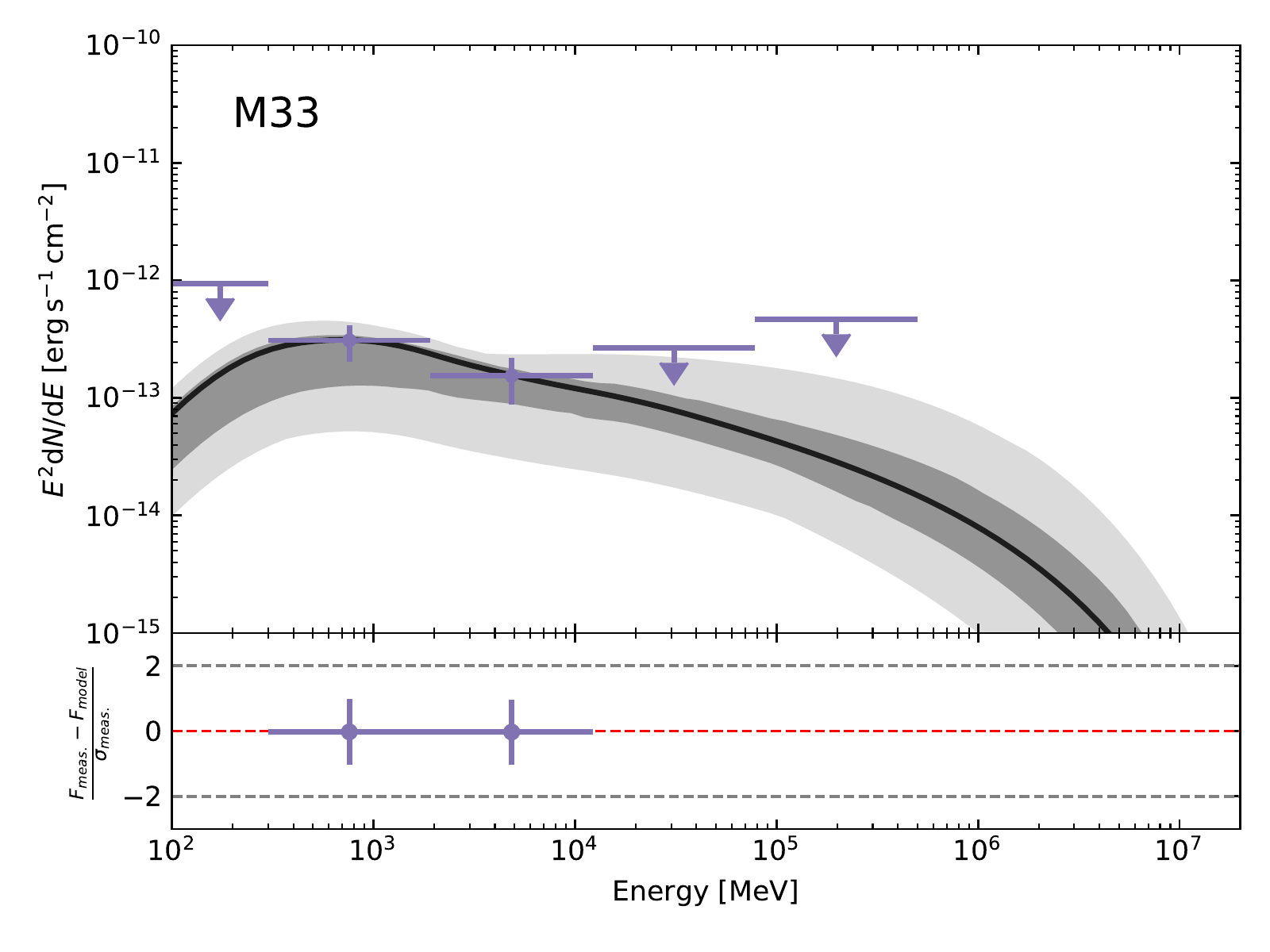}
\caption{Modeling of the GeV band SEDs of SBGs and SFGs.
The black solid lines are the best-fit results of the one-zone model. 
Purple data points are from the \textit{Fermi}-LAT observation. The dark gray and light gray represent the 1$\sigma$ and 3$\sigma$ confidence intervals of the model fit, respectively. The panels below each figure are the residual figure of the best-fit result. 
}
\label{Fig2}
\end{figure*}

Diffusive shock acceleration (DSA), as a predominant acceleration  mechanism, is generally used to explain high-energy particle acceleration at SNR shocks of up to approximately 100 TeV or even higher \citep{Aharonian2007,Aharonian2011,Morlino2012a}.
Therefore, here the range of protonic energy was selected from 0.1 GeV to 0.5 PeV \citep[e.g.,][]{Abdalla2018}. 
When the protonic energy distributions of the SBGs and SFGs satisfy (1), we used \textbf{NAIMA} to obtain the best-fit results of the SEDs of all sources, as shown in Figure \ref{Fig2}. The related best-fit parameters are  given in Table \ref{Tab2}, where the protons   energy budget $W_{\rm p}$ was calculated to be above 290 MeV, which is considered as the pion production threshold \citep{Abdalla2018}. 
Except for the hardening spectral components from NGC 1068 and NGC 4945,  
we found that the one-zone model with ECPL can explain the SEDs of most SBGs and SFGs. 
However,  
we found that the first bins of SEDs of NGC 1068, M82, NGC 253, Arp 299, and SMC cannot be well explained; we  assumed that the first bins may be from the contributions of the Bremsstrahlung emission and inverse Compton scattering of the primary or secondary electrons of CRs \citep[e.g.,][]{Tang2014}.


\begin{table*}
\renewcommand\arraystretch{1.2}
\renewcommand\tabcolsep{3.0pt} 
\caption{Spectral Fitting Parameters }

\begin{center}
\begin{tabular}{cccccccccc}
\hline
Source Name & Distance& Redshift & Gas Density & $\alpha$ & $E_{\rm cutoff}$ & $ W_{\rm p}$ & $\chi_{\rm red}^{2}$/$N_{\rm dof}$ & $N_{\rm SNR}$ & References \\
 \hline
    & (mpc) & & (cm$^{-3}$) & & (TeV) & (erg) &  & 10$^{4}$ &\\ 
 \hline
 \multicolumn{8}{c}{The Starburst Galaxy} \\ 
 \hline     
 NGC 253   & 2.5 & 0.0009 & 250 & 2.38$_{-0.02}^{+0.02}$ &  159.36$_{-40.78}^{+38.48}$ &$2.23_{-0.33}^{+0.40}\times 10^{53}$ & $\frac{2*7.31}{15-4}=1.33$ & 2.23 & (9), (2), (19) \\
 
 M82       & 3.4 &0.0007  & 175 & 2.38$_{-0.02}^{+0.03}$ &  $166.41_{-27.15}^{+18.13}$ &$9.57_{-0.72}^{+0.64}\times 10^{53}$ &$\frac{2*9.93}{12-4}=2.48$ & 9.57 & (9), (2),(18) \\
 
  NGC 2146  & 15.2&0.003 & 10 & 2.13$_{-0.11}^{+ 0.12}$ & 39.01$_{-3.32}^{+2.72}$  &$3.10_{-0.43}^{+0.88}\times 10^{55}$ &$\frac{2*0.71}{6-4}=0.71$ & 310.18 & (3), (2), (20)\\
  
 NGC 4945  & 3.7 &0.001 & 1000  & 2.50$_{-0.05}^{+0.06}$ & 0.78$_{-0.37}^{+0.26}$  & $2.79_{-0.76}^{+0.63}\times 10^{53}$ &$\frac{2*5.12}{10-4}=1.71$ & 2.79 & (10), (2), (19)\\
 NGC 1068  & 16.7& 0.004& 120 & 2.93$_{-0.04}^{+0.03}$ & 10.49$_{-1.13}^{+1.06}$   & 6.72$_{-0.73}^{+0.83}\times 10^{55}$ &$\frac{2*9.37}{9-4}=3.75$ & 672.33 & (7), (2),(19)\\
  Arp 220   & 77 & 0.018 & 3500 &2.84$_{-0.08}^{+0.06}$ & 9.47$_{-1.49}^{+1.49}$ &$2.27_{-0.71}^{+0.53}\times 10^{55}$ & 
  $\frac{2*0.90}{7-4}=0.60$ & 227.21 & (9), (11), (16)\\ 
   \hline
  \multicolumn{8}{c}{Contains the likely contribution of AGNs} \\ 
   \hline
 Circinus  & 4.2 &0.001 & 500  & 2.36$_{-0.10}^{+0.11}$  & 1.80$_{-0.50}^{+0.57}$   &$2.63_{-0.44}^{+0.92}\times 10^{53}$  & $\frac{2*0.28}{6-4}=0.28$ & 2.63 & (13), (15), (17)\\ 
 NGC 3424  & 26.2 & 0.005 & 10 & 2.15$_{-0.34}^{+0.18}$ & 3.51$_{-0.91}^{+0.49}$ &8.97$_{-0.56}^{+0.92}\times 10^{55}$  &$\frac{2*0.70}{6-4}=0.70$ & 897.22 & (11),-,(21)\\ 
 Arp 299   & 47.74&0.001 & 70 & 2.05$_{-0.68}^{+0.64}$ & 0.25$_{-0.12}^{+0.15}$ &$2.86_{-0.62}^{+0.75}\times 10^{55}$  & $\frac{2*0.70}{5-4}=1.40$ & 286.46 & (12), (11), (22)\\
 \hline 
 \multicolumn{8}{c}{The Star-forming Galaxy} \\
 \hline
 M31       & 0.78 & -0.001 & 0.6 &  2.11$_{-0.18}^{+0.20}$  & $0.03_{-0.008}^{+0.007}$  &4.19$_{-1.33}^{+1.73}\times 10^{54}$ &$\frac{2*0.01}{6-4}=0.01$ & 41.93 & (4), (1), (23)\\ 
 M33       & 0.93 & -0.0006 & 100 & 2.49$_{-0.08}^{+0.05}$  & 9.64$_{-1.39}^{+1.94}$  &1.93$_{-0.50}^{+0.41}\times 10^{52}$  &$\frac{2*0.0001}{5-4}=0.0002$ & 0.19 & (6), (5), (23) \\
 SMC       & 0.06 &0.0005 & 0.2 & 2.52$_{-0.05}^{+0.08}$ & 0.16$_{-0.04}^{+0.03}$  &$1.38_{-0.62}^{+0.65}\times 10^{54}$ &$\frac{2*31.50}{10-4}=10.50$ & 13.81 & (8), (1), (23)\\
 LMC       & 0.05 &0.0009 & 2 & 2.58$_{-0.03}^{+0.03}$  & 2.19$_{-0.46}^{+0.32}$   &2.86$_{-0.62}^{+0.69}\times 10^{53}$ &$\frac{2*14.73}{10-4}=4.91$ & 2.86 & (14), (1), (23)\\ 
 \hline
\end{tabular}
\end{center} 
 \flushleft 
Note: The last column is the references of distance, gas density, and redshift of each source, respectively. The column of $N_{\rm SNR}$ is SNR's number inside each source.
Here we assumed that the gas density of NGC 3424 was 10 cm$^{-3}$ since its spectrum and photon flux of the global fit  were similar to those of NGC 2146. 
(1) \citep{Ackermann2012a}, (2) \citep{Gao2004}, (3) \citep{Greve2006}, (4) \citep{Kavanagh2020}, (5) \citep{Karachentsev2017}, (6) \citep{Kramer2020}, (7) \citep{Lamastra2019}, (8) \citep{Lopez2018}, (9) \citep{Peretti2019}, (10) \citep{Roy2010}, (11) \citep{Sanders2003}, (12) \citep{Sargent1987}, (13) \citep{Wang2018}, (14) \citep{Tang2017}, (15) \citep{Tully2009}, (16)\citep{Ahn2012}, (17)\citep{Skrutskie2006}, (18)\citep{Abazajian2009},(19)\citep{Meyer2004}, (20)\citep{Falco1999}, (21)\citep{van2016},(22)\citep{Izotova1999}, (23)\citep{McConnachie2012},(24)\citep{Paturel2002}
\label{Tab2}
\end{table*}

\subsection{Two-zone Model fit for NGC 1068 and NGC 4945}

\begin{figure*}[!h]
\includegraphics[width=73mm]{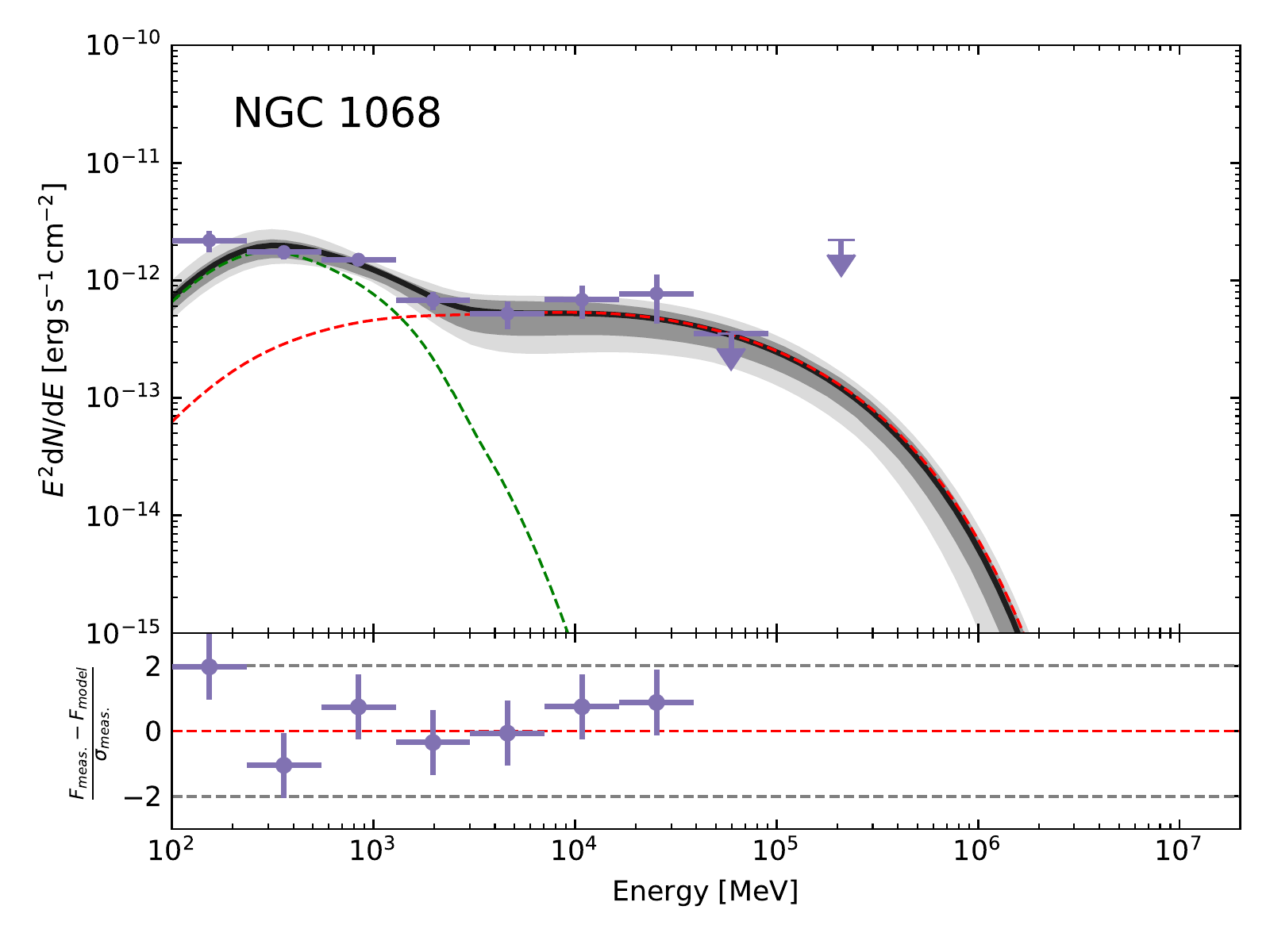}
\includegraphics[width=73mm]{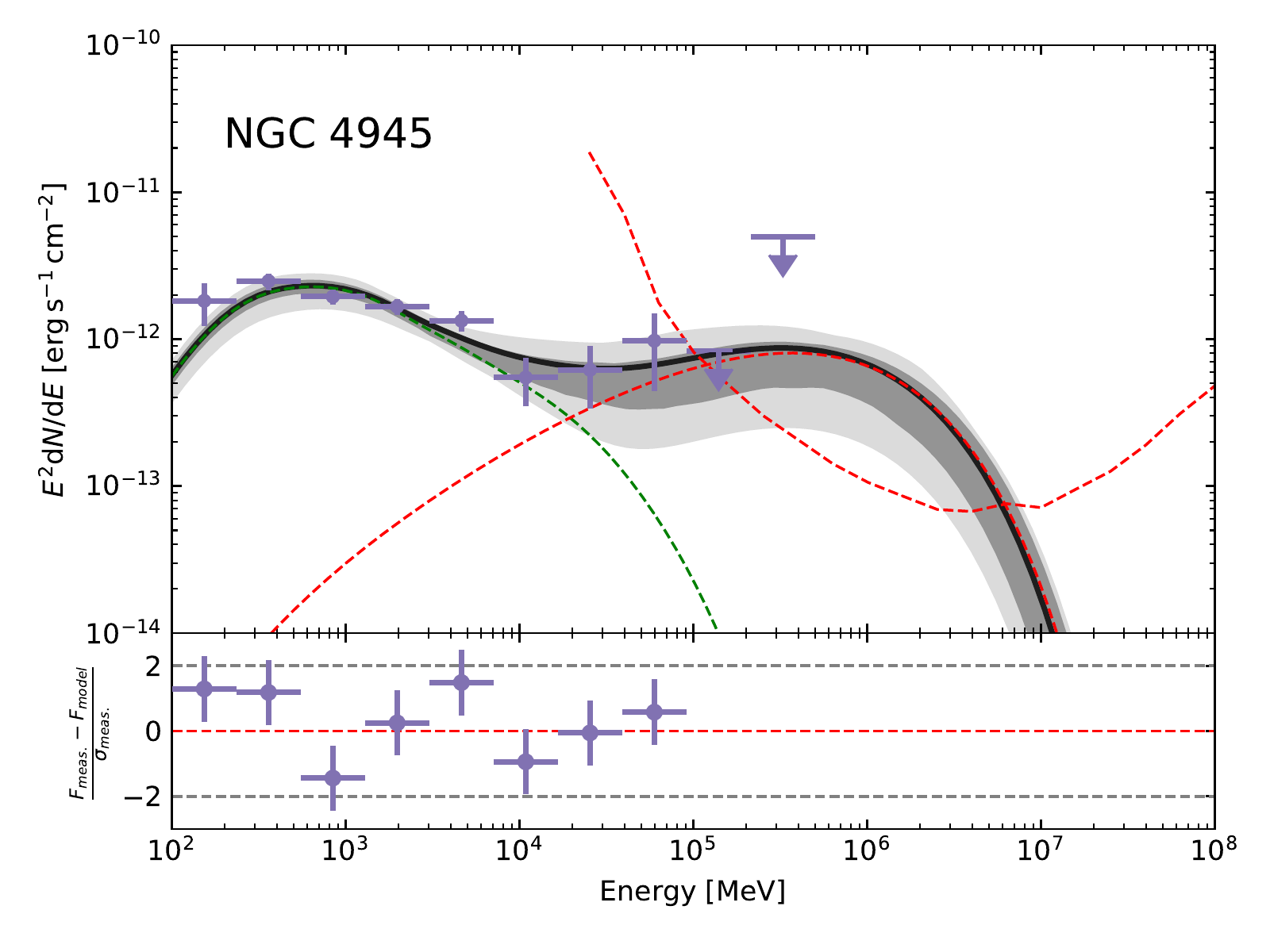}
\includegraphics[width=73mm]{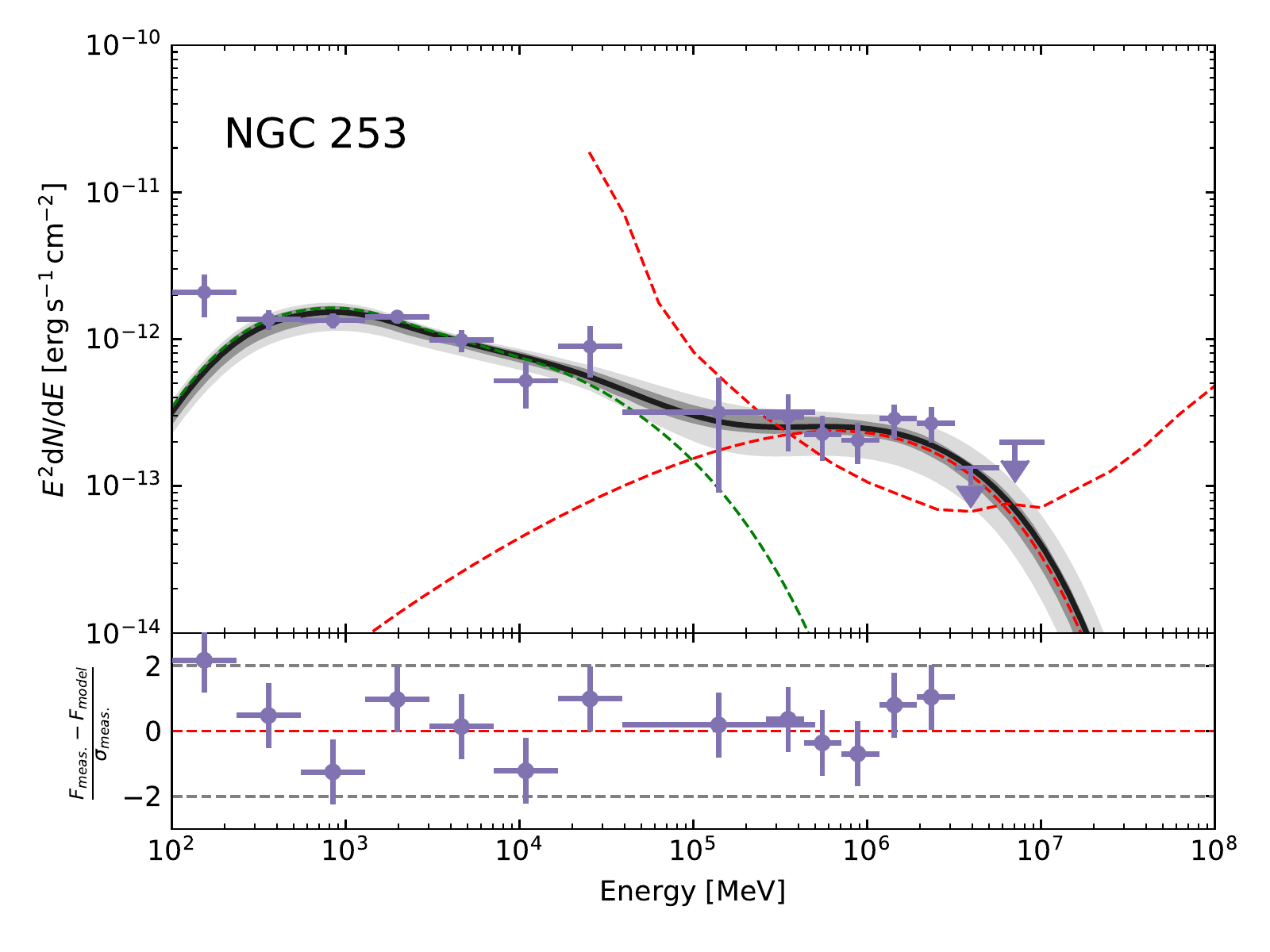}
\includegraphics[width=73mm]{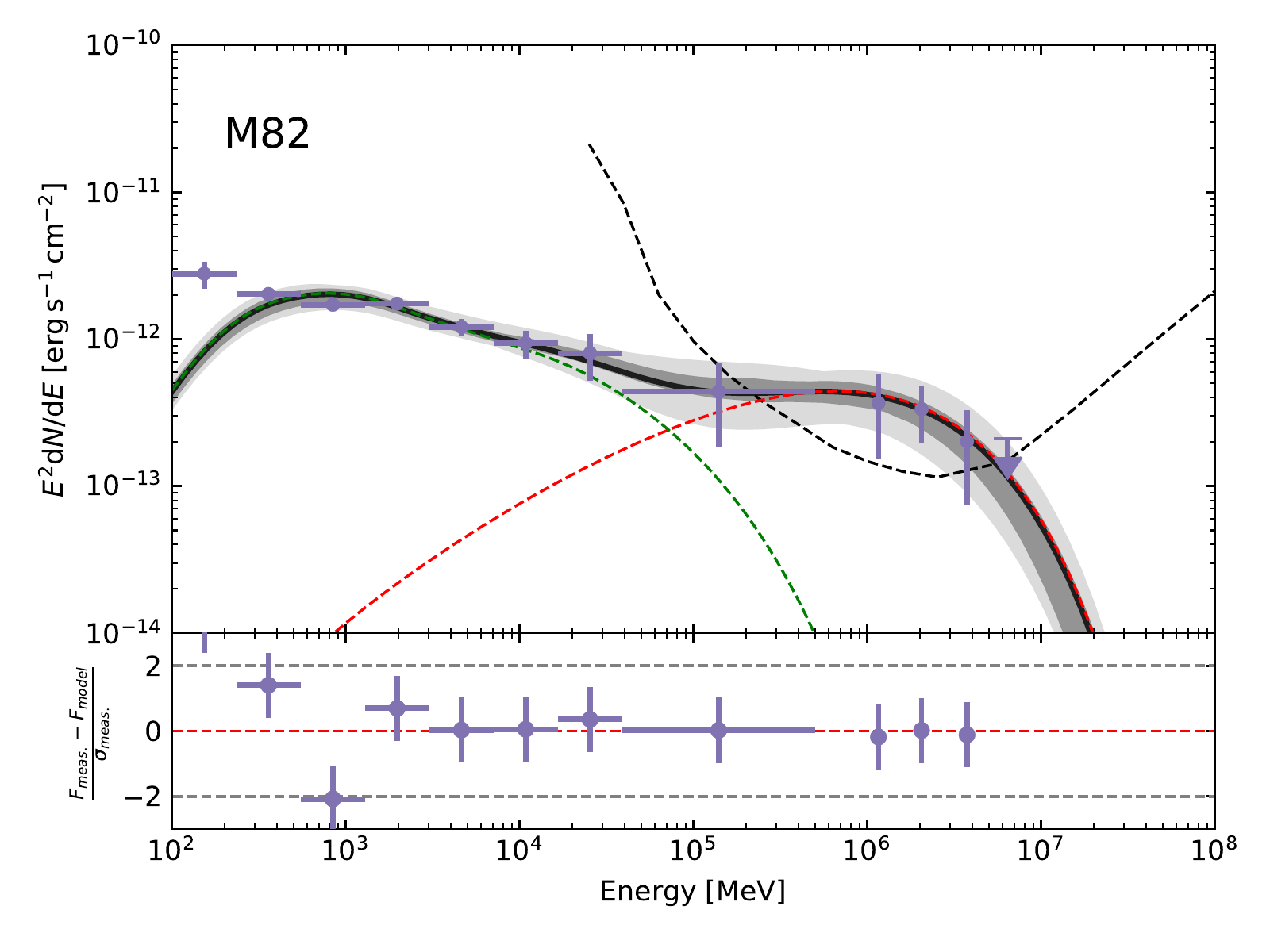}
\caption{These panels show the best-fit results of the SEDs of NGC 1068, NGC 4945, NGC 253, and M82 from the two-zone model. The red and green dashed lines represent  hadronic component 1 and component 2, respectively. The black solid line represents the sum of the two hadronic components. Other descriptions are as same as Figure \ref{Fig1} and Figure \ref{Fig2}. 
}
\label{Fig3}
\end{figure*}

As can be seen in Figure \ref{Fig2}, we found that the one-zone model did not explain the hardening spectral components from NGC 1068 and NGC 4945. 
Therefore, we considered a two-zone model, 
which assumes that they have two kinds of different protonic compositions from   the two different zones. 
All protons inside the two zones satisfy the ECPL energy distribution. 
The two zones contributed to the soft and hard spectral compositions in the 0.1 to 500 GeV band. 
As shown in Figure \ref{Fig3}, by fitting their SEDs, we found that this two-zone model can better explain the hardening components of the SEDs of NGC 1068 and NGC 4945 than the one-zone model. Their best-fit results are given in Table \ref{Tab3}.

\begin{table*}[!h]
\caption{The Best-fit Parameters of The Two-zone Models}
\begin{center}
\begin{tabular}{lccccccc}
  \hline\noalign{\smallskip}
    \hline\noalign{\smallskip}
  Model Parameter        &  gas density    & $E_{0}$ & $\alpha$               & $E_{\rm cutoff}$       & $W_{\rm p}$  & $\chi^{2}/N_{dof}$   \\
                         &  (cm$^{-3}$) & (TeV)                        &                    &  (TeV) & (erg)   &  \\
  \hline\noalign{\smallskip}

  \multicolumn{6}{c}{NGC 1068}  &  $\frac{3.4*2}{9-8}=6.8$  \\
    \noalign{\smallskip}\hline 
  Hadronic component1 (red dashed)    & 120 & 20 & $1.89_{-0.08}^{+0.05}$  & $0.80_{-0.12}^{+0.07}$ & $ 5.04_{-0.70}^{+0.66} \times 10^{54}$   &     \\
                     
    \noalign{\smallskip}\hline
 Hadronic component2 (green dashed)   & 120 & 1 & $3.33_{-0.05}^{+0.03}$  & $ 0.007_{-0.001}^{+0.001}$ & $ 2.08_{-0.41}^{+0.45} \times 10^{56}$   &     \\
   \noalign{\smallskip}\hline
   
     \multicolumn{6}{c}{NGC 4945} &  $\frac{2*4.37}{10-8}=4.37$  \\
       \noalign{\smallskip}\hline
  Hadronic component1 (red dashed)  & 1000 &  20 & $1.12_{-0.19}^{+ 0.13}$  & $5.20_{-0.75}^{+ 1.13}$ & $ 1.87_{-0.45}^{+0.35} \times 10^{52}$   &     \\
    \hline\noalign{\smallskip}
  Hadronic component2 (green dashed)   & 1000 & 1 & $2.50_{-0.06}^{+0.04}$  & $ 0.18_{-0.01}^{+0.03}$ & $ 3.30_{-0.41}^{+0.58} \times 10^{53}$   &     \\
  \noalign{\smallskip}\hline  
  
       \multicolumn{6}{c}{NGC 253} &  $\frac{2*6.3}{15-8}=1.8$  \\
       \noalign{\smallskip}\hline
  Hadronic component1 (red dashed)  & 250 &  6 & $1.24_{-0.12}^{+ 0.11}$  & $13.20_{-1.30}^{+ 2.22}$ & $ 1.04_{-0.10}^{+0.14} \times 10^{52}$   &     \\
    \hline\noalign{\smallskip}
  Hadronic component2 (green dashed)   & 250 & 1 & $2.33_{-0.04}^{+ 0.03}$  & $ 0.65_{-0.13}^{+0.15}$ & $ 2.55_{-0.88}^{+0.73} \times 10^{53}$   &     \\
  \noalign{\smallskip}\hline  

       \multicolumn{6}{c}{M82} &  $\frac{2*9.5}{12-8}=4.75$  \\
       \noalign{\smallskip}\hline
  Hadronic component1 (red dashed)  & 175 &  3 & $1.24_{-0.19}^{+ 0.18}$  & $11.20_{-1.50}^{+1.24}$ & $ 3.24_{-0.25}^{+0.36} \times 10^{52}$   &     \\
    \hline\noalign{\smallskip}   
  Hadronic component2 (green dashed)   & 175 & 1 & $2.38_{-0.04}^{+0.04}$  & $ 0.73_{-0.09}^{+0.04}$ & $ 7.88_{-0.48}^{+0.82} \times 10^{53}$   &     \\
  \noalign{\smallskip}\hline

\end{tabular}
\end{center}
\label{Tab3}
\end{table*}

\section{Discussion}
\subsection{Spectral Feature} 
All sources are classified into three categories, including SBG with the likely contribution from internal AGN (SBG-AGN), SBG without AGN (SBG), and SFR.
For SBG-AGN, there are three objects: Circinus \citep{Guo2019}, NGC 3424 \citep{Peng2019}, and Arp 299 \citep{Xi2020}. 
Here, we first analyzed the spectral features of the three types of sources.
For SBGs, we excluded NGC 4945 and NGC 1068 because they cannot be well explained by the one-zone model; we  excluded Arp 220 because its fifth energy bin was associated with a higher flux than the fourth energy. 
 By calculating the average values of the spectral index of NGC 253, M82, and NGC 2146, we found that the value was approximately 2.30. 
  For SBG-AGNs and SFRs, their average values of the spectral index were 2.19 and 2.43, respectively. 
These results suggest that the spectral features of SBGs and SBG-AGNs are hard, and there is not much difference between them; thus,  we suggest that the internal hidden AGNs from SBG-AGNs may not have a significant contribution to the GeV high-energy $\gamma$-ray emission.

The spectrum of M31 is currently hard, though there are only two data points with TS value $>$ 4. 
We suggest that its particles may be in a primary acceleration stage because the process of particle acceleration from internal SNRs generally requires a long timescale from hundreds to thousands of years after the SNR explosion \citep{Yuan2018}. 
In the current \textit{Fermi}-LAT observation period of approximately 11 years, its particles may not be accelerated to a high-energy band, which makes the $E_{\rm cutoff}$ value of the SED of M31  appear at about 0.03 TeV. 
On the other hand, if its inner particles are at the stage of the late evolution, this also results in a low $E_{\rm cutoff}$ in the  spectrum for the radiative cooling effect \citep[e.g.,][]{Brantseg2013,Cox1972,Tang2013,Zeng2019,Xiang2021c}.
Similarly, the particles within M33 with a low $E_{\rm cutoff}$ of  approximately 9.64 TeV are likely to be in an early or late evolution stage. 


\subsection{Likely Protonic Acceleration Limit within SBGs} 

Owing to the lack of TeV data for most SBGs and SFRs thus far, there is no definitive conclusion about the possible range of $E_{\rm cutoff}$. Currently, there are only two TeV SBGs, including NGC 253 and M82. Based on the simulation of the one-zone model, we found their values of $E_{\rm cutoff}$ to be approximately 159.36 TeV and 166.41 TeV, respectively. 
 These results indicate that their internal protons can be accelerated to the order of 10$^{2}$ TeV, which is close to the particle acceleration limit of current SNRs in the Milky Way  \citep{Aharonian2007,Aharonian2011,Morlino2012a}.

\citet{Berezhko2003} analyzed the X-ray data of SN 1006 from Chandra observations and confirmed the magnetic field amplification, which can effectively accelerate nuclear CRs in SNRs to TeV or PeV levels   \citep{Tang2013,Lin2019}. 
Similar to the SNRs in the Milky Way, if the SBGs contain  numerous SNRs and the energy of CR particles in these SNRs can be accelerated to the 10$^{2}$ TeV level, then there should be a strong magnetic field amplification near the shock surface in SNRs \citep{Lin2019}. 
 If the evolution of CR particles in NGC 253 and M82 is  similar to that of SNRs in the Milky Way, then the magnetic field amplification is likely to be an important mechanism for the acceleration of particles inside them to the 10$^{2}$ TeV level. 
To derive the limitations of particle acceleration, more high-energy data are expected in the future (e.g., continuous \textit{Fermi}-LAT observations).

According to the previous works \citep[e.g,][]{Zeng2019,Xiang2021a,Xiang2021c}, in the Milky Way, for the hadronic scenario, each SNR needs the average value of $W_{\rm p}$ to be approximately 10$^{49}$ erg in the high-energy band. Based on the value of $W_{\rm p}$, we estimated the number of SNR in each source in Table \ref{Tab2}.


\subsection{Two-zone SED for NGC 1068 and NGC 4945} 

There are three SBGs, including NGC 3424, Circinus, and Arp 299, which are considered to have the contribution of AGNs in the GeV band \citep{Guo2019,Peng2019,Xi2020}. The main reason is that some variability is found in their light curves of approximately ten years.
Thus, the contribution of AGNs cannot be ignored for these SBGs.
The correlation of the X-ray and $\gamma$-ray emission of NGC 4945 indicated that the $\gamma$-ray emission was likely dominated by AGN \citep{Wojaczynski2017}. Moreover, the $\gamma$-ray emission of NGC 1068 was believed to be likely from its AGN activity, considering that it exceeded the expectation of the star-forming process  \citep{Lenain2010,Ackermann2012a,Eichmann2016}.  However, The GeV flare is an important feature of AGN and has not been observed from NGC 4945 and NGC 1068 thus far \citep{Ackermann2012a,Peng2019,Ajello2020}. 
In addition, they well conformed to the two well-known luminosity relations; one is that of the total IR (8-1000 $\mu$m) and the $\gamma$-ray luminosities; the other is that of the $\gamma$-ray and 1.4 GHz radio continuum luminosities. These results suggest that their $\gamma$-ray emissions may originate from the star-forming process \citep{Yun2001,Thompson2007, Ackermann2012a, Guo2019, Xi2020, Ajello2020}. 
The previous research results provide strong evidence to support the  hadronic origin of SNR for NGC 4945 and NGC 1068 in this analysis.


For SNRs, \citet{Uchiyama2007} found the variable X-ray filaments (or knots) within the region of RX J1713.7-3946 with Suzaku and Chandra. These filaments themselves could contribute significantly to the $\gamma$-ray emission from RX J1713.7-3946 \citep{Finke2012}.
This result implies that the one-zone fit is inadequate to explain the overall SEDs of SNRs in the Milky Way. Therefore, \citet{Finke2012} 
considered a multi-zone scenario to well explain the multi-band SED of RX J1713.7-3946. 
Furthermore, \citet{Zhang2016} used a two-zone model to well fit the broadband spectrum of RX J1713.7-3946. 
In their model, the first zone was the cavity wall of SNR  with the dense matter at the cavity boundary; the second zone was inside the SNR. 
\citet{Lu2020} considered a two-zone model with different diffusion processes from the extended region of SNR and the internal pulsar wind nebula region to explain the broadband SED of SNR G21.5-0.9. Similarly, \citet{Xiang2021b} also used a simple two-zone model,  irrespective of the acceleration process of CR particles, to well explain overall SED from composite SNR G327.1+1.1.
If the GeV emissions of NGC 1068 and NGC 4945 are two-zone collective contributions from their internal SNRs,  we found that the two-zone model can well explain the hardening phenomena of NGC 1068 and NGC 4945 at the end of their GeV spectra. 

Since NGC 253 and M82 have GeV and TeV data points, we also used the two-zone model to fit their SEDs to verify that the model is generally useful for explaining the high-energy origin of current SBGs. The best-fit results of NGC 253 and M82 are shown in Figure \ref{Fig3}. 
We find that the two-zone model can also explain the GeV to TeV SEDs of the two SBGs.
In Table \ref{Tab3}, for the four SBGs, we can see that the spectra of hadronic component 1 are harder than those of hadronic component 2;  the values of $W_{\rm p}$ of the former are one or two orders of magnitude lower than those of the latter.  
In the future, more observations with high precision in the high-energy band are required to confirm  the two-zone hypothesis. 


\section{Conclusion}

1. We regenerated the SEDs of nine SBGs and four SFGs using Fermitools. Owing to the uncertainty of the acceleration mechanism inside the SBGs and SFGs, we used the simple one-zone model provided by \textbf{NAIMA} to explain their GeV spectra. Moreover, we considered the two-zone model to better explain the hardening components from the spectra of NGC 1068 and NGC 4945 than the one-zone scenario. 

2. Classifying all sources into three categories, we found that the spectral features of SBGs and SBG-AGNs were not significantly different from the average spectral indexes of 2.30 and 2.19, respectively. The SFG spectra were relatively soft, with an average spectral index of 2.43.

3. Analyzing the $E_{\rm cutoff}$ of these sources, we found that the $E_{\rm cutoff}$ of NGC 253 and M82 attained the order of 10$^{2}$ TeV, which indicates that the internal protons can be accelerated to the order of 10$^{2}$ TeV in the one-zone scenario.

4. For the protons energy budget $W_{\rm p}$, we found that the average values of $W_{\rm p}$ from SBGs and SBGs-AGNs were approximately 10$^{54}$ erg, and that of SFGs was approximately 10$^{53}$ erg. 


\section{Acknowledgements}

For this work, we sincerely thank the referee for his/her invaluable
comments and gratefully acknowledge Li Zhang, Pengfei Zhang, Jianeng Zhou, Fangkun Peng, Xian Hou, Zhongxiang Wang, Shenbang Yang, Yuliang Xin, Dahai Yan, Liangliang Ren, and Jun Fang for their generous help. We also appreciate the support for this work from 
National Key R\&D Program of China under grant No. 2018YFA0404204, the National Natural Science Foundation of China (NSFC U1931113, U1738211, U193110119), the Foundations of Yunnan Province (2018IC059, 2018FY001(-003)), the Scientific research fund of Yunnan Education Department (2020Y0039).


\end{document}